\documentclass[pra,twocolumn,aps,floatfix,10pt,superscriptaddress]{revtex4}
\usepackage{amsmath}
\usepackage{amsfonts}
\usepackage{amssymb}
\usepackage{graphicx}
\usepackage{ucs}
\usepackage[utf8x]{inputenc}
\usepackage{psfrag,layout}
\usepackage{upgreek}
\usepackage{subfigure}
\usepackage{bm}
\usepackage{color}
\usepackage{xfrac}
\usepackage{hyperref}

\newcommand{\varone}{(t,\vec{r})}
\newcommand{\vartw}{(\tau,\vec{x})}
\newcommand{\varthr}{(s,\vec{x})}

\def\Xint#1{\mathchoice
	{\XXint\displaystyle\textstyle{#1}}%
	{\XXint\textstyle\scriptstyle{#1}}%
	{\XXint\scriptstyle\scriptscriptstyle{#1}}%
	{\XXint\scriptscriptstyle\scriptscriptstyle{#1}}%
	\!\int}
\def\XXint#1#2#3{{\setbox0=\hbox{$#1{#2#3}{\int}$ }
		\vcenter{\hbox{$#2#3$ }}\kern-.6\wd0}}

\def\dashint{\Xint-}

\providecommand{\norm}[1]{\lVert#1\rVert}

\renewcommand{\Im}{\operatorname{Im}}

\renewcommand{\vec}[1]{\mathbf{#1}}
\newcommand{\diff}{\mathop{}\!\mathrm{d}}

\newcommand{\dadv}[1]{\frac{\mathrm{D} #1}{\mathrm{D} t} }

\newcommand\mydots{\hbox to 1em{.\hss.\hss.}}

\begin{document}

\title{Quantum hydrodynamics for supersolid crystals and quasicrystals}

\author{Vili Heinonen}
\affiliation{
	Department of Mathematics, Massachusetts Institute of Technology,
	77 Massachusetts Avenue, Cambridge, Massachusetts 02139-4307, USA
	}
\email{vili@mit.edu}
\author{Keaton J. Burns}
\affiliation{
  Department of Physics, Massachusetts Institute of Technology, Cambridge, MA 02139-4307, USA
	}
\author{J\"orn Dunkel}
\affiliation{
	Department of Mathematics, Massachusetts Institute of Technology,
	77 Massachusetts Avenue, Cambridge, Massachusetts 02139-4307, USA
	}
\email{dunkel@mit.edu}

\begin{abstract}
Supersolids are theoretically predicted quantum states that break the continuous rotational and translational symmetries of liquids while preserving superfluid transport properties.
Over the last decade, much progress has been made in understanding and characterizing supersolid phases through numerical simulations for specific interaction potentials. 
The formulation of an analytically tractable framework for generic interactions still poses theoretical challenges. By going beyond the usually considered quadratic truncations, we derive a systematic higher-order generalization of the Gross-Pitaevskii mean field model in conceptual similarity with the Swift-Hohenberg theory of  pattern formation. We demonstrate the tractability of this broadly applicable approach by determining the ground state phase diagram and  the dispersion relations for the supersolid lattice vibrations in terms of the potential parameters. Our analytical predictions agree well with numerical results from direct hydrodynamic simulations and  earlier quantum Monte-Carlo studies. The underlying framework is universal and can be extended to anisotropic pair potentials with complex Fourier-space structure.
\end{abstract}


\maketitle

\section{Introduction}
Supersolids are superfluids in which the local density spontaneously arranges in a state of crystalline order. The existence of supersolid quantum states was proposed in the late 1960s  by Andreev, Lifshitz and Chester~\cite{Andreev1969,Chester1970} but the realization in the lab has proven extremely difficult~\cite{Boninsegni2012}.  Recent experimental breakthroughs in the control of ultracold quantum gases~\cite{Kadau2016,Leonard2017,Li2017,Chomaz2018} suggest that it may indeed be possible to design quantum matter that combines dissipationless flow with the intrinsic stiffness of solids.  Important theoretical insights into the expected properties of supersolids and experimental candidate systems have come from numerical mean field calculations and quantum Monte-Carlo (qMC) simulations for specific particle interaction potentials~\cite{Pomeau1993,PomeauYves;Rica1994,Macri2013,Cinti2014}. What is still lacking, however, is a general analytically tractable framework that allows the simultaneous characterization of whole classes of potentials as well as the direct computation of ground state phase diagrams and dispersion relations for supersolid lattice vibrations in terms of the relevant potential parameters.
\par
To help overcome such conceptual and practical limitations, we introduce here a generalization of the classical Gross-Pitaveskii (GP) mean field model~\cite{Gross1961,Pitaevskii1961} by drawing  guidance from the Swift-Hohenberg theory~\cite{1977SwiftHohenberg}, which uses fourth-order partial differential equations (PDEs) to describe pattern formation in Rayleigh-B\'enard convection. Our approach is motivated by the successful application of higher-than-second-order PDE models to describe classical solidification phenomena~\cite{Elder2002,Elder2004,Emmerich2012}, electrostatic correlations in concentrated electrolytes and ionic liquids~\cite{2011Bazant,2012Storey}, nonuniform FFLO superconductors~\cite{1994BK_PLA}, symmetry breaking in elastic surface crystals~\cite{Stoop2015}, and active fluids~\cite{2012Wensink,Bratanov08122015,Dunkel2013_PRL}.  Whereas higher PDEs are often postulated as effective phenomenological descriptions of systems with crystalline or quasicrystalline order~\cite{2009Cross}, it turns out that such equations can be derived directly within the established GP framework. The resulting mean field theory yields analytical predictions that agree with direct hydrodynamic and recent qMC simulations~\cite{Henkel2012,Cinti2014} and specify the experimental conditions for realizing periodic supersolids and  coexistence phases, as well as supersolid states exhibiting quasicrystalline symmetry (Sec.~\ref{s:quasi-SS}).

\section{Mean Field Theory}
As in the classical GP theory~\cite{Gross1961,Pitaevskii1961}, we assume that a system of spinless particles can be described by a complex scalar field $\Psi(t,\vec  x)$ and that quantum fluctuations about the mean density $n(t,\vec x)= |\Psi|^2 $ are negligible. Considering an isotropic pairwise interaction potential $u(|\vec  x-\vec  x'|)$, the total potential energy density is given by a spatial convolution integral which can be expressed as a sum over Fourier mode contributions $\propto \hat{u} |\hat{n}|^2$, where overhats denote Fourier transforms (see App.~\ref{app:derivation_energy} for a detailed derivation). If $u$ is isotropic with finite moments, then its Fourier expansion can be written as $ \hat{u}(\vec  k) =  \sum_{j=0}^{\infty} g_{2j} k^{2j}$ \cite{2007Elder_PRB}, where $k=|\vec  k|$ is the modulus of the wave vector, yielding the potential energy density $\frac{1}{2}  n \sum_{j=0}^{\infty}(-1)^{j} g_{2j}  (\nabla^2)^j n$ with $\nabla^2$ denoting the spatial Laplacian. The constant Fourier coefficients $g_{2j}$ encode the spatial structure of the potential. Variation of the total energy functional with respect to $\Psi$, yields the generalized GP equation (App.~\ref{app:derivation_energy})
\begin{equation}
i \hbar \partial_t \Psi  = \left[-\frac{\hbar^2}{2 m} \nabla^2 
+ \left(\sum_{j=0}^{\infty}(-1)^{j} g_{2j}  (\nabla^2)^j \left| \Psi \right|^2\right) \right]\Psi .
\label{eq:gross-pitaevskii}
\end{equation}
The classical GP model, corresponding to repulsive point interactions $u=g_0\delta(\vec  x-\vec  x')$ is recovered for $g_0>0$ and $g_{2j}=0$ otherwise. 
The authors of Ref.~\cite{Veksler2014} studied the case $g_0,g_2>0$ and $g_{2j\ge 4}=0$, keeping partial information about long-range ($k\to 0$) hydrodynamic interactions by effectively adding a surface energy term  $ \propto |\nabla n|^2 $ to the energy density. However, as we shall see shortly, to describe supersolid states, 
the long-wavelength expansion has to be carried out at least to order $k^4$. 

\begin{figure*}[t]
	\includegraphics[width = 1.95\columnwidth]{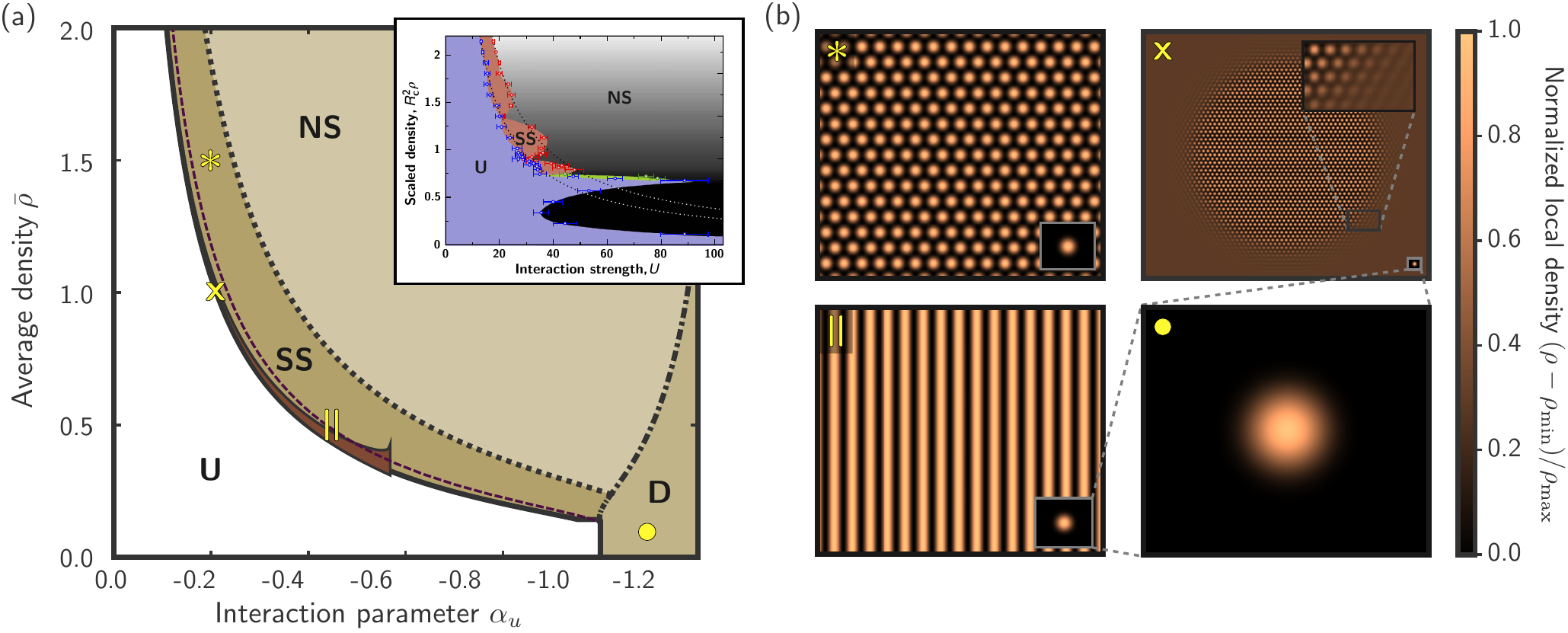}
	\caption{(a) Phase diagram showing the ground state structure of the potential energy functional $U[\rho]$ in Eq.~\eqref{eq:potential_energy}, calculated analytically using a one-mode approximation. The inset shows the phase diagram for an interacting Rydberg-dressed BEC obtained from qMC simulations, adapted with permission from Fig.~2 in Ref.~\cite{Cinti2014}. (b) Examples of numerically computed minima of $U[\rho]$ for parameters indicated by the symbols $ (\bm{\ast} ,  \bm{\mathsf{x}} , \bm{\parallel},  \bullet ) $ in panel (a). The uniform superfluid state (U) is stable against perturbations below the thin dashed red line.  The  hexagonal supersolid (SS, $\bm{\ast}$; Movie~1~\cite{supplemental}) phase has a lower energy than the metastable supersolid stripe phase (SS, $ \bm{\parallel} $).  The first-order transition between U-phase and SS-phase supports a narrow coexistence region (dark brown)   with the uniform subphase having a lower density ($\bm{\mathsf{x}}$).   In the NS state,  the one-mode minimization of $U$ yields locally negative density solutions, signaling failure of the approximation (see also Fig.~2).  In the droplet phase  ($ \bullet $) no real-valued one-mode solutions exist.  The single-droplet solution (D, $ \bullet $) is shown as an inset for scale in the other panels $ (\bm{\ast} ,  \bm{\mathsf{x}} , \bm{\parallel})$.
}
	\label{pic:phase_diagram}
\end{figure*}

\par
To show this explicitly, it is convenient to express the complex dynamics~\eqref{eq:gross-pitaevskii} in real Madelung form~\cite{2013Berloff_PNAS}.
Writing $\Psi = \sqrt{n} \exp(i S)$ and defining the irrotational velocity field $ \vec{v} = (\hbar/m)\nabla S $,  
Eq.~\eqref{eq:gross-pitaevskii} can be recast as the hydrodynamic equations~(App.~\ref{app:hydrodynamic_formulation}, see also \cite{Wyatt2005quantum})
\begin{align}
\label{eq:massevolution-main}
\partial_t n  &= -\nabla \cdot \left( n  \vec{v} \right)
\\
\label{eq:velocityevolution-main}
m (\partial_t + \vec{v}\cdot \nabla) \vec{v}  &=  -\nabla (\delta U/\delta n) 
\end{align}
with the effective potential energy $U[n]$  to order~$\mathcal{O}(k^4)$
\begin{equation}
\label{e:quantum-potential}
U=\!
\int\! \diff\vec{x}\left[ \vphantom{\frac{\left( \nabla^2 n \right)^2}{2}}
\frac{\hbar^2}{8m}\frac{|\nabla n|^2}{n}+  \frac{g_0}{2} n^2 + \frac{g_2}{2}|\nabla n |^2
+ \frac{g_4}{2}\! \left( \nabla^2 n \right)^2
\right].
\end{equation}
The first term is the kinetic quantum potential~\cite{2013Berloff_PNAS} and the kinetic energy is $K=({m}/{2})\int d\vec{x}\, n\vec{v}^2$.  For non-leaky boundary conditions, Eqs.~\eqref{eq:massevolution-main} and \eqref{eq:velocityevolution-main} conserve the total particle number $N=\int d\vec{x}\, n$ and energy $E[n,\vec{v}] = K[n,\vec{v}] + U[n]$. 
The dynamics defined by Eqs.~\eqref{eq:massevolution-main}-\eqref{e:quantum-potential} is Hamiltonian in terms of the momentum density  $m n \vec{v}=\hbar n \nabla S$, thereby differing, for example, from generalized Ginzburg-Landau theories for nonuniform FFLO superconductors~\cite{1994BK_PLA}. 
Local minima of $E[n,\vec{v}]$ have zero flow, $\vec{v}\equiv 0$, and hence the corresponding density fields must minimize $U[n]$. 
Assuming short range repulsion $g_0>0$,  we see that uniform constant-density solutions minimize $U[n]$ when $g_2 \geq 0$ and $g_4=0$~(App.~\ref{app:uniform_ground_state}); this case was studied in Ref.~\cite{Veksler2014}. 
If, however,  we consider the more general class of pair interaction potentials with $g_2\gtrless 0$ then short-wavelength stability at order $k^4$ requires that $g_4>0$. This situation arises, for example, for the two-dimensional (2D) Rydberg-dressed Bose-Einstein condensate (BEC) studied in Ref.~\cite{Henkel2012}, which has $g_0=0.189\hbar^2/m$, $g_2=-0.113\,(\hbar^2/m)\,\mu$m$^2$  and $g_4 = 0.016 \, (\hbar^2/m)\,\mu$m$^4$~(App.~\ref{app:RB_parameters}). Whereas for $g_2>0$ roton-minima are absent as in the classical GP theory~\cite{Pomeau1993},  supersolid ground state solutions~\cite{Pomeau1993,PomeauYves;Rica1994} of Eq.~\eqref{e:quantum-potential} can exist when $g_2<0$, as we will see shortly.

\par

To determine the ground state phase diagram of $U[n]$ with $g_2<0$, it is convenient to define the characteristic wave number \mbox{$ q_0 = \sqrt{-g_2/(2 g_4)}>0$}, time scale $t_0=m/(\hbar q_0^2)$, and  energy scale  $\epsilon_0= \hbar^2 q_0^2/m $. Focusing on the 2D case  and adopting $(q_0^{-1},t_0,\epsilon_0)$ as units from now on, we can rewrite Eq.~\eqref{e:quantum-potential} as
\begin{equation}
\label{eq:potential_energy}
	U[\rho] = \frac{1}{2} \int  \diff{\vec{x}}\left[\frac{|\nabla \rho|^2}{4\rho} +  \alpha_u \rho^2 
	+ \rho \left(
	1 + \nabla^2 
	\right)^2 \rho\right]
\end{equation}
where $\rho =  mn (g_4 q_0^2 / \hbar^2)$  is the rescaled number density and   
\begin{equation}
\label{e:alpha}
\alpha_u =  \frac{4 g_0 g_4}{g_2^2} -1
\end{equation}
the interaction parameter (App.~\ref{app:hydrodynamic_formulation}). On an infinite domain, the internal energy $U[\rho]$ is completely parameterized by $\alpha_u$ and 
the average density $ \bar{\rho} = \int \rho \diff \vec{x}/ \int \diff\vec{x}$. For the Rydberg-dressed BEC in Ref.~\cite{Henkel2012},  
which has $q_0 =  1.87\, \mu \text{m}^{-1} $ corresponding to a hexagonal lattice spacing of $ 3.88\, \mu$m,
one finds $ \alpha_u = -0.043 $ and $\bar{\rho} \approx 9.4$ (App.~\ref{app:RB_parameters}).

\section{Ground State Structure}
An advantageous aspect of the above framework is that the ground state structure of $U[\rho]$ in the $(\alpha_u, \bar{\rho})$-phase plane can be 
explored both analytically and numerically in a fairly straightforward manner (Fig.~\ref{pic:phase_diagram}). Standard linear stability analysis (App.~\ref{app:linear_stability})
for uniform constant-density solutions predicts a symmetry breaking transition at  $\alpha_u=-1$ when $\bar{\rho} \le 1/8$ and 
\begin{equation}
\alpha_u=\frac{1-16\bar\rho}{64\bar\rho^2} \qquad\text{when}\qquad  \bar\rho > 1/8
\end{equation}
indicated by the thin dashed line in Fig.~\ref{pic:phase_diagram}(a). Refined analytical estimates for the ground state phases can be obtained by considering the Fourier ansatz $ \rho = \bar{\rho} +  \sum_{j} \phi_j \exp{(i \vec{q}_j \cdot \vec{x})} $. The pattern forming operator \mbox{$ (1 + \nabla^2)^2 $} penalizes modes with $ |\vec{q}_j| \neq 1 $, suggesting that single-wavelength expansions can yield useful approximations for the ground state solutions.  Conceptually similar studies for classical phase field models~\cite{Emmerich2012} imply that 2D and 1D close-packing structures realizing  hexagonal and stripe patterns are promising candidates.  The one-mode approximation ansatz for a hexagonal lattice reads 
\begin{equation}
\label{eq:hexagonal_one-mode}
	\rho = \bar{\rho} + \phi_0 \sum_{j=1}^{3} [\exp{(i  \vec{q}_j \cdot \vec{x})} + \text{c.c.}],
\end{equation}
where the lattice vectors $\vec{q}_j $ form the ``first star'' with $ \vec{q}_i \cdot \vec{q}_j = q^2 $, if $ i=j $ and $-q^2/2 $ otherwise (c.c. denotes complex conjugate terms). Similarly, the stripe phase is defined by $ \rho = \bar{\rho} + [\phi_0 \exp{(i q x_1)} + \text{c.c.}]$. These trial functions have to be minimized with respect to the amplitudes $ \phi_0 $ and the magnitudes $q$~of the reciprocal lattice vectors, which can be done analytically (App.~\ref{app:one-mode_approximation}). 
Our analytical calculations predict four distinct pure ground state types that can be identified as uniform (U), supersolid (SS), normal solid-like (NS) and droplet (D), and also a narrow U/SS coexistence phase via Maxwell construction [dark brown domain in Fig.~\ref{pic:phase_diagram}(a)]. These naming conventions are directly adopted from Ref.~\cite{Cinti2014}. As always, mean field predictions should be supplemented with other methods to properly characterize the supersolid-NS transition to ensure that the wave functions are localized in the NS state \cite{svistunov2015}. Yet, the comparison with the qMC simulations of  Ref.~\cite{Cinti2014} suggests that  our mean field results capture essential aspects of their numerical results, see inset of Fig.~\ref{pic:phase_diagram}(a).
\par
The uniform U-phase has constant density $ \rho = \bar{\rho}$. The supersolid SS-phase is distinguished through the existence of non-zero amplitude solutions  $\phi_0$ that yield strictly positive periodic density patterns $\rho >0$.  By contrast, in the NS state, the one-mode density field can become locally negative, signaling a breakdown of the one-mode approximation.   In the droplet D-phase, no real-valued density solutions exist within the one-mode approximation. 
\par
\begin{figure}
	\includegraphics[width =0.9\columnwidth]{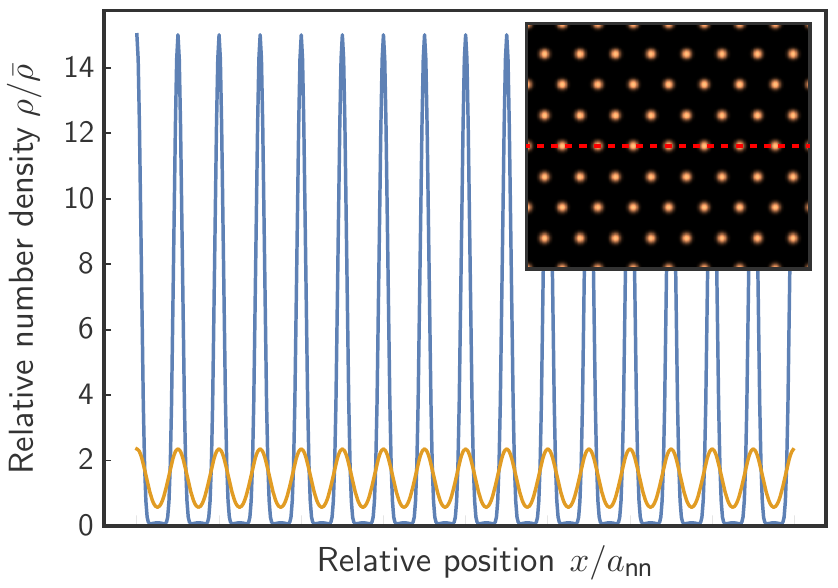}
	\caption{Difference between supersolid (SS) and normal solid-like (NS) states. 1D cross-sections of the density fields of two numerically determined ground states with same average density $ \bar{\rho} = 0.4 $  but $ \alpha_u= -0.484$ (SS, orange) and $-0.60$ (NS, blue), respectively.  In both cases, the positions are rescaled with the nearest neighbor distance $ a_\text{nn} = 4\pi/\sqrt{3}q $ to compensate the different lattice spacings. Inset: full 2D density field for $\alpha_u = -0.6 $; dashed red line indicates the 1D cross-section.
	}
	\label{pic:one-mode_breakdown}
\end{figure}
To test the analytical predictions and explore the validity of the underlying one-mode approximation in detail, we performed a numerical ground state search (App.~\ref{app:numerical_methods}) for various parameter pairs $(\bar{\rho},\alpha_u)$. Representative examples of numerically found states are shown in Figs.~\ref{pic:phase_diagram}(b) and~\ref{pic:one-mode_breakdown} and generally agree well the analytical predictions. The hexagonal supersolid at $\bar{\rho}=1.5 , \alpha_u = -0.2 $ [$ \bm{\ast} $ in Fig.~\ref{pic:phase_diagram}(b)] is indeed dominated by a single mode, with the second highest mode being $\sim 20$ times smaller.  Our analytical one-mode estimates suggest that the supersolid stripe states have a higher energy than hexagonal states, with the energy difference going to zero as one approaches the U-SS phase boundary [solid line in Fig.~\ref{pic:phase_diagram}(a)].  This opens the possibility that systems at non-zero kinetic energy or in a strained geometry might arrange in a stripe configuration similar to the metastable state shown at $ \bar{\rho} = 0.5 $, $ \alpha_u = -0.45 $  [$\bm{\parallel} $ in Fig.~\ref{pic:phase_diagram}(b)]. The numerical solution for $ \alpha_u = -0.2178 $, $ \bar{\rho}=1.0 $ [$ \bm{\mathsf{x}} $ in Fig.~\ref{pic:phase_diagram}(b)] confirms  the analytically predicted coexistence of uniform and supersolid phases, suggesting that the one-mode approximation places the coexistence region accurately in the phase diagram. 
For coexistence to be observable in experiments, the energy of the uniform and supersolid bulk regions has to be significantly larger than that of the interface, which requires a sufficiently large system size. At high $ \bar{\rho} $, the coexistence gap closes, approaching the asymptote $ \alpha_u = -0.22/\bar{\rho}$  of the phase transition.
\par
Our simulations show that the one-mode approximation describes the ground state structure accurately down to average densities $ \bar{\rho} \gtrsim 0.4 $ near the uniform--supersolid phase transition. If $\bar{\rho}$ and/or $ \alpha_u$ are reduced further, higher modes will no longer be negligible.  The difference between a supersolid one-mode ground state and a normal solid multi-mode solution at the same average density $ \bar{\rho}=0.4$ is shown in Fig.~\ref{pic:one-mode_breakdown}. The density profile for the NS state resembles a collection of Gaussian peaks, which typically have a significantly larger separation than the peaks of a one-mode solution at similar $\bar{\rho}$. Finally, the domain $ \alpha_u < -1 $ corresponds to a negative GP parameter $ g_0 < 0 $, thus representing attractive contact interactions. The numerically determined ground state at $ \bar{\rho} = 0.1 $, $ \alpha_u = -1.1 $ [$ \bullet $ in Fig.~\ref{pic:phase_diagram}(b)] realizes a single droplet with an approximately Gaussian density profile, qualitatively similar to recent experimental observations of quantum droplet formation  in dilute $^{39}$K BECs~\cite{Cabrera2017}.

\section{Dynamics}
\begin{figure}
	\includegraphics[width =0.9\columnwidth]{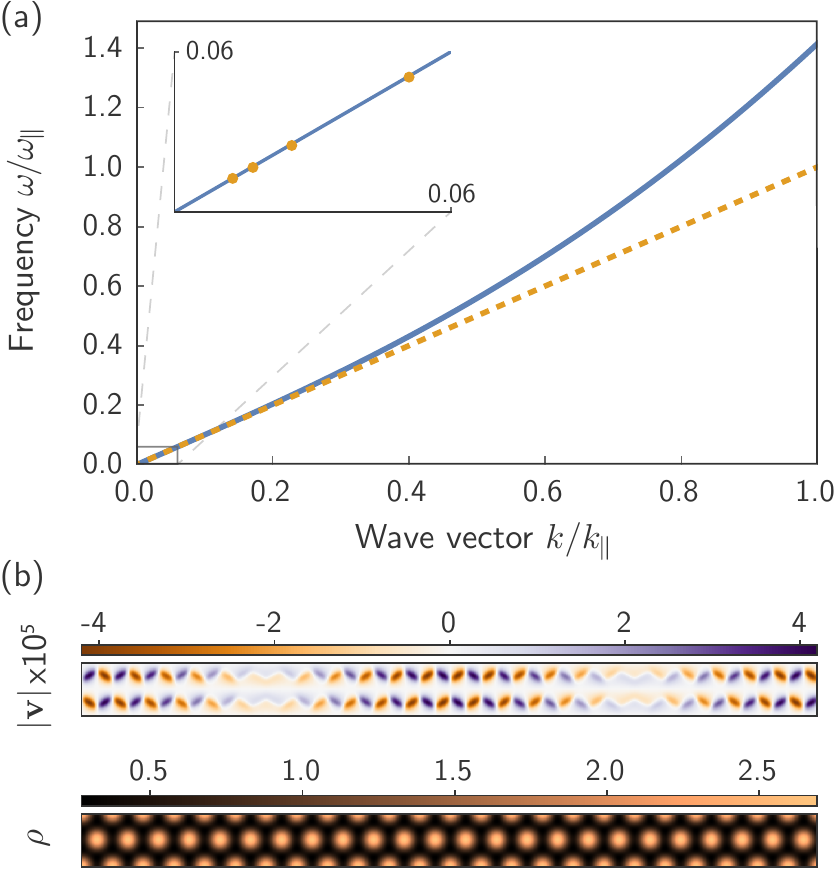}
	\caption{(a) The analytically calculated nonlinear dispersion relation (blue) for plane waves becomes asymptotically linear for small $k$-values (dashed orange). See text for the scaling factors $ \omega_\parallel $ and $ k_\parallel $. Inset: The analytical prediction (blue line) agrees well with numerical results (symbols) obtained by solving the full nonlinear dynamical system described by Eqs.~\eqref{eq:massevolution-main} and \eqref{eq:velocityevolution-main} for $ \bar{\rho} = 1.0 $ and $ \alpha_u = -0.24 $. (b)~Representative snapshot from a numerical simulation showing the magnitude of the velocity field and the density field. See also Movie 2 (SM \cite{supplemental}).
	}
	\label{pic:dispersion_relation}
\end{figure}

We next describe how the above framework can be used to obtain predictions for the sound wave dynamics in a supersolid (see App.~\ref{app:wave_equation} for details of calculations). The supersolid phase breaks the continuous translational symmetry and supports lattice vibrations. These vibrations can be studied analytically close to the uniform--supersolid phase transition, where the one-mode approximation for the density $ \rho $ is accurate.  Near the phase transition the local deviation  of $ \rho $ from its mean value $\bar{\rho}$ is relatively small, and one can Taylor-expand the nonlinear quantum potential  in $ \rho $  around $ \bar{\rho} $. Adopting this approximation, we now consider a change of coordinates $ \vec{x} \to \vec{x} - \vec{u}(t,\vec{x})$, where  $ \vec{u} $ is a displacement field in the Eulerian frame $ \vec{x} $.  Since we are interested in hydrodynamic long-wavelength sound excitations, we may assume that  the displacement field varies over a length scale significantly larger than the spacing between the hexagonal density peaks $|\nabla \vec u|\ll 1$. Inserting the one-mode ansatz~\eqref{eq:hexagonal_one-mode}  into Eq.~\eqref{eq:potential_energy},  and keeping only the leading terms  in $ \vec{u}$, one obtains an energy functional $U[\vec u]$ that is quadratic in the displacement field (App.~\ref{app:wave_equation}).  The approximative dynamics  of $\vec u$ is found from  Eqs.~\eqref{eq:massevolution-main} and~\eqref{eq:velocityevolution-main} by noting~\cite{Heinonen2016_2} that, for small displacements, $ \partial_t \vec{u} = \vec{v} $ and $  \partial_t \vec{v} = - \nabla \delta U/\delta \rho = -(1/\bar{\rho} )\delta U/\delta \vec{u}$ hold. This gives the linear equation
\begin{equation}
\begin{split}
\partial_t^2 \vec{u} &= \frac{3 \phi_0^2 q^2}{\bar{\rho}} \left[
\frac{1-\phi_0/\bar{\rho} + 5 \phi_0^2 / \bar{\rho}^2}{4 \bar{\rho}}
 \nabla^2 \vec{u} \right. \\& \left. 
 \quad
+   \left(
 3 q^2 - 2 
 \right)  \nabla^2 \vec{u} 
+ 2 q^2 \nabla (\nabla \cdot \vec{u}) 
-\frac{2}{3} \nabla^4 \vec{u} \vphantom{\frac{1-\phi_0/\bar{\rho} + 5 \phi_0^2 / \bar{\rho}^2}{4 \bar{\rho}}}
\right],
\end{split}
\label{eq:wave_equation}
\end{equation}
which is solved by a plane wave $ \vec{u}_0 \exp{[i( \vec{k} \cdot \vec{x}-\omega t)]}$.  Since the field $ \vec{v}\propto \nabla S $ describes an irrotational potential flow with  $ \nabla \times \vec{v} = 0 $, only longitudinal modes are allowed. Inserting the plane wave ansatz in Eq.~\eqref{eq:wave_equation} yields the nonlinear dispersion relation $ 	\omega/\omega_\parallel  = (k/k_\parallel) \sqrt{1+(k/k_\parallel)^2} $, with $ k_\parallel^2 = \frac{3}{2} \left\{[(5 \phi_0^2/\bar{\rho} ^2-\phi_0/\bar{\rho} +1)/4 \bar{\rho}] - 2 + 5 q^2  \right\} $ and $ \omega_\parallel^2 =  2 q^2 \phi_0 ^2 k_\parallel^4/\bar{\rho}$, which is shown as a blue line in Fig.~\ref{pic:dispersion_relation}(a). For supersolids described by  the one-mode approximation, one has $ q \approx 1 $ and $ \phi_0/\bar{\rho} <1/3$ implying $k/k_{||}\ll 1$. This suggests that vibrations with wavelengths larger than the lattice constant generally exhibit a linear dispersion, and  that the speed of sound is simply given by~$ c_\parallel = \omega_\parallel/k_\parallel $  [dashed orange curve in Fig.~\ref{pic:dispersion_relation}(a)]. 
 
\par
The analytical prediction derived from the linear Eq.~\eqref{eq:wave_equation} agrees well with the numerical dispersion results obtained by simulating the full nonlinear Eqs.~\eqref{eq:massevolution-main} and \eqref{eq:velocityevolution-main} with the open-source spectral code Dedalus~\cite{dedalus2017}, see inset of Fig.~\ref{pic:dispersion_relation}(a). In our simulations, we analyzed a longitudinal mode in a periodic box, for four different box sizes in the direction of the wave, using $ \bar{\rho} = 1.0 $ and $ \alpha_u = -0.24 $ (App.~\ref{app:numerical_methods}). For these parameters, one finds $ \phi_0 = 0.266$ and $q = 0.934$ by minimizing the energy $ U $ of the supersolid state with the one-mode approximation. 
Representative simulation snapshots showing the magnitude of the velocity field and the density can be seen in Fig.~\ref{pic:dispersion_relation}(b); see also Movie 2 (SM \cite{supplemental}). The low-$k$ mode is visible as an enveloping modulation of the velocity field.  The sound speed was measured in the simulations by estimating  $w_\parallel $ and $ k_\parallel $ from a linear fit, yielding $c_\parallel= 0.6994 \pm 0.0015 $, in excellent agreement with the theoretically predicted value~$\omega_\parallel/k_\parallel = 0.6995 $. Thus, although waves can scatter from inhomogeneities of the density field in the fully nonlinear system, the long-wavelength dynamics is accurately captured by the linearized theory.

\section{Superfluid quasicrystals}
\label{s:quasi-SS}
\begin{figure}
	\centering
	\includegraphics[width =0.95\columnwidth]{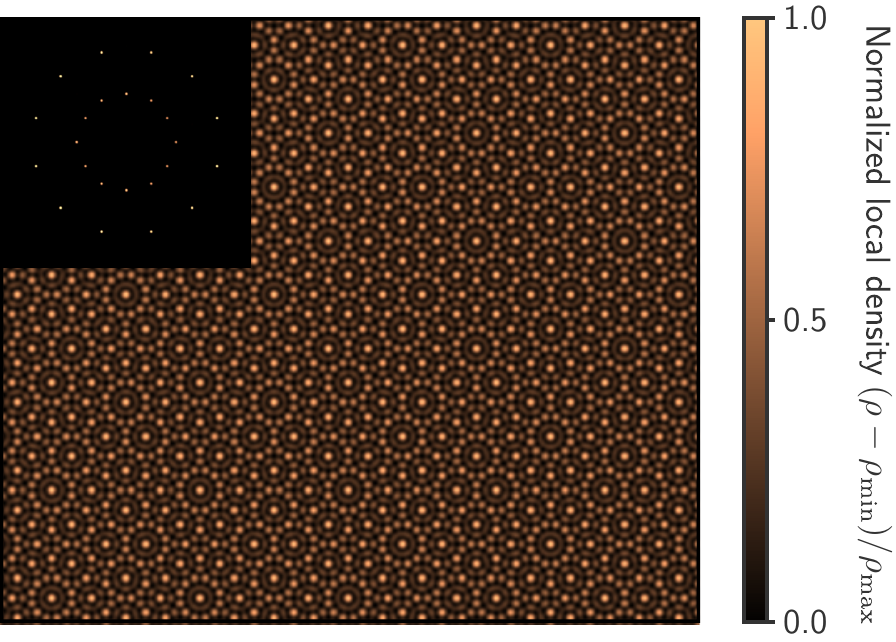}
	\caption{Superfluid ground state exhibiting quasicrystalline order, as predicted by Eq.~\eqref{eq:gross-pitaevskii} for an interaction potential $ u $ having relevant terms up to 8th order in the expansion~\eqref{eq:higher_order_expansion}. System parameters:  $ \bar{\rho} = 1.0 $, $ \alpha_u = -0.9 $, $ b_1 = 0.09 $, $ b_2 = 0 $, $ a_1 = 0.982 $, $ a_2 = 2 \cos{(\pi/12)} $.  Here $ \rho_\text{min} \approx 0.395 $ and $ \rho_{\text{max}} \approx 2.969 $. The minimum density is well above 0 suggesting that this state is superfluid.   
		The inset shows the absolute value of the Fourier transformed density, $ |\hat{\rho}| $, indicating a 12-fold rotational symmetry with two rings in the reciprocal lattice. See also Movie 3 (SM \cite{supplemental}.
	}
	\label{pic:quasicrystal}
\end{figure}

The supersolid phase in Fig.~\ref{pic:phase_diagram} is caused by the pattern forming $ \alpha_u (1+\nabla^2)^2 $-term in the internal energy $U$. This fourth-order contribution makes plane waves with wave number $k_0=|k| = 1 $ energetically favorable, giving rise to a hexagonal pattern. More complex supersolid structures can be expected in systems where the Fourier transformed interaction potential $ \hat{u} $ possesses multiple local minima $ k_i $ where $ \hat{u}(k_i) < 0 $. To demonstrate that this is indeed the case, let us  replace $(1+\nabla^2)^2$ by
 \begin{equation}
 	 [b_1 + (a_1^2+\nabla^2)^2][b_2 + (a_2^2+\nabla^2)^2],
 	\label{eq:higher_order_expansion}
 \end{equation}
which corresponds to keeping terms up to order $j=4$ in the generalized GP equation~\eqref{eq:gross-pitaevskii}. The resulting 8th-order operator in Eq.~\eqref{eq:higher_order_expansion} is energetically bounded from below and accounts for an asymmetry in the local potential energy minima $ k_i $ through the parameters~$ b_i $. Note that $ a_i $ and $k_i $  are in general not equal anymore when $b_i\ne 0$.  With a larger number of physically relevant length scales at play, systems can attain a wider range of ground state structures \cite{Mkhonta2013,Lifshits}.  To illustrate this, we computed the ground state of Eq.~\eqref{eq:gross-pitaevskii} for the expansion~\eqref{eq:higher_order_expansion} with parameters  $ \bar{\rho} = 1.0 $, $ \alpha_u = -0.9 $, $ b_1 = 0.09 $, $ b_2 = 0 $, $ a_1 = 0.982 $, $ a_2 = 2 \cos{(\pi/12)} $.  As evident from the corresponding mean field density $ \rho $ and its absolute Fourier transform $ |\hat{\rho}| $ in Fig.~\ref{pic:quasicrystal}, the ground state in this case is a \emph{quasicrystalline} superfluid state with 12-fold rotational symmetry; see also Movie 3 (SM \cite{supplemental}).
\par
In this context, we mention that earlier discussions~\cite{Gopalakrishnan2013,Hou2018} of superfluid quasicrystalline states considered spin and pseudospin interactions  with only one relevant length scale. By contrast, the quasicrystalline ground state in Fig.~\ref{pic:quasicrystal} forms due to the competition between the two length scales set by the local minima of the interatomic potential~$ \hat{u}$. While this mechanism is reminiscent of classical quasicrystal pattern formation \cite{Mkhonta2013,Lifshits}, the mean field quantum systems discussed here differ from their classical counterparts through the presence of the quantum potential term which penalizes high $ k $ modes.
\bigskip

\section{Conclusion}
To conclude, we introduced and studied a higher-order generalization of the classical GP mean field theory that accounts for the structure of pair interactions through the relevant Fourier coefficients of the underlying potential.   The resulting hydrodynamic equations share many conceptual similarities with classical Swift-Hohenberg-type pattern formation models~\cite{1977SwiftHohenberg}, for which a wide range of advanced mathematical analysis tools exists~\cite{Lloyd08,Thiele,Emmerich2012}. The generalized GP equation~\eqref{eq:gross-pitaevskii} allows to transfer these techniques directly to quantum systems. By focusing on the fourth-order case,  we obtained analytic predictions for ground state phase diagrams and low-energy excitations that agree well with direct numerical simulations. With regard to experiments, our results  suggest that the coexistence phase at the first-order superfluid-supersolid transition may be the most promising regime for observing supersolids. We expect this mean field prediction to be robust as the analytically derived uniform-supersolid phase transition curve  (solid line in Fig.~\ref{pic:phase_diagram}) agrees well recent qMC simulations~\cite{Cinti2014}  that account for beyond mean field effects  [Fig.~\ref{pic:phase_diagram}(a) inset].

When the interaction parameter $ \alpha_u$ in Eq.~\eqref{e:alpha} becomes too small, a transition to a normal solid state is expected as quantum fluctuations will likely destroy phase coherence in the low-density domains. 
However, grain boundaries of polycrystalline normal solid could still show superfluid behavior \cite{Sasaki1098,Pollet2007} and can be studied using the theory presented here. 
Interestingly, our numerical ground state analysis suggests that at low average densities a quantum droplet can be stabilized by the kinetic quantum potential without requiring higher-order nonlinearities to correct for quantum fluctuations~\cite{Cabrera2017,Petrov2015}. 
From a conceptual perspective,  the generalized GP framework appears well suited for future extensions: 
By calculating the quasicrystalline ground state (Fig.~\ref{pic:quasicrystal} and Movie~3~\cite{supplemental}), we have already demonstrated how to extend the approach to systems with multiple relevant length scales. 
The theory can be refined by including Lee-Huang-Yang~\cite{LeeYang_1,Lee1957} corrections $\propto \rho^{5/2}$ that penalize peaked densities. 
Generalizations to anisotropic potentials and multi-component systems seem both feasible and experimentally relevant~\cite{Kadau2016,Cabrera2017,PhysRevA.97.023619}. 
We thus hope the above discussion can provide useful guidance for future efforts in these and other directions. 

\begin{acknowledgments}
This work was supported by the Finnish Cultural Foundation (V.H.) and an Edmund F. Kelly Research Award (J.D.).
\end{acknowledgments}

\appendix

\section{Derivation of the mean field energy}\label{app:derivation_energy}
\subsection{Mean-field reduction of many-particle Schr\"odinger equation}
For completeness, we present the derivation of the generalized GP equation~\eqref{eq:gross-pitaevskii} from a many-particle Schr\"odinger equation with a pair potential $ u(\vec{r}_i,\vec{r}_j) $. Our starting point is a bosonic $ N $-particle quantum system, described by a wave function $ \Psi_N (t,\vec{r}_1,\mydots \, ,\vec{r}_N) $ that satisfies the exchange symmetry
\begin{equation}
\Psi_{N}(t, \mydots \,,\vec{r}_i,\mydots \,, \vec{r}_j,\mydots ) = +\Psi(t, \mydots \,, \vec{r}_j, \mydots\,, \vec{r}_i, \mydots ).
\end{equation}
The dynamics of the system is governed by the $ N $-particle Schr\"odinger equation
\begin{equation}
	i \hbar \partial_t \Psi_N = \hat{H}_{N} \Psi_N,
\label{eq:Ndynamics}
\end{equation}
where the $ N $-particle Hamiltonian is defined by
\begin{equation}
\begin{split}
	 \hat{H}_{N}  &= -\alpha \sum_i \nabla_{\vec{r}_i}^2 +\sum_i v_{\text{ext}}(\vec{r}_i) \\ &
	 +\frac{1}{2}\sum_{i=1}^N\sum_{j\ne i}^N\ u(\vec{r}_i,\vec{r}_j)
\end{split}
\label{eq:NHamiltonian}
\end{equation}
with $ \alpha = \hbar^2/(2m) $. Equation~\eqref{eq:Ndynamics} can be written as a variation of the expectation value of $ \hat{H}_N $: 
\begin{equation}
\begin{split}
i \hbar \partial_t \Psi_N = 
	\frac{\delta E}{\delta \Psi_{N}^*},
\end{split}
\end{equation}
where
\begin{equation}
	E_{\Psi_{N}}[\Psi_N,\Psi_N^*] := \left\langle \hat{H}_{N} \right\rangle = \int \diff \vec{r}^N \Psi_N^* \hat{H}_{N} \Psi_N.
\label{eq:Nenergy}
\end{equation}
The key step in the derivation of the generalized GP equation is the mean field approximation 
\begin{equation}
\label{eq:meanfieldansatz}
\begin{split}
	\Psi_N &(t,\vec{r}_1,...,\vec{r}_N) = \\
	&\frac{1}{N!}\sum_{\pi \in S_N}  \psi(t,\vec{r}_{\pi(1)})\psi(t,\vec{r}_{\pi(2)}) \mydots \psi(t,\vec{r}_{\pi(N)}),
\end{split}
\end{equation}
where the sum is over the permutation group $ S_N $ of indices $ 1, \mydots\, N $. This approximation is equivalent to assuming that there are no entangled particles in the system. It also implies that the one-particle probability distributions $ \psi^*(\vec{r}_i) \psi (\vec{r}_i) $ are independent and identical. 
\par
We can use Eq.~\eqref{eq:meanfieldansatz} to calculate the mean field approximation of the energy. The symmetric form of the mean field ansatz allows for relabeling indices in the evaluation of the Hamiltonian $ H_{N} $ giving 
\begin{equation}
\begin{split}
E_{\Psi} &=  \int \diff \vec{r} \left(
-\alpha \Psi^* \nabla^2  \Psi + |\Psi|^2 v_{\text{ext}}
\right) + \frac{N (N-1)}{2 N^2} \\&
\times \int \diff \vec{r}_1 \diff \vec{r}_2 \left[
\Psi^* (\vec{r}_{1}) \Psi (\vec{r}_1) u(\vec{r}_{1},\vec{r}_{2}) \Psi^* (\vec{r}_{2}) \Psi (\vec{r}_{2})
\right],
\end{split}
\end{equation}
Here $ \Psi = \sqrt{N} \psi  $ so $ |\Psi|^2 $ becomes the number density $ n $. 
For large $ N $,  one has $ N (N-1)/N^2 \approx 1 $. 
\par
To derive a dynamical equation for the one-particle wave function, we evaluate the expectation value of the time evolution operator. Multiplying  Eq.~\eqref{eq:Ndynamics} by $ \Psi_{N}^* $ and integrating over the spatial coordinates gives
\begin{equation}
\begin{split}
	\left \langle i \hbar \partial_t \right\rangle  
	&= \int \diff \vec{r}^N \Psi_{N}^* i \hbar \partial_t \Psi_{N} \\
	&= \int \diff \vec{r}^N \Psi_{N}^* i \hbar \sum_{j=1}^N \frac{\partial \Psi_{N}}{\partial \Psi (\vec{r}_j) } \partial_t \Psi (\vec{r}_j) \\
	&= \frac{1}{\sqrt{N}} \int \diff \vec{r}^N \Psi_{N}^* i \hbar \sum_{j=1}^N \frac{ \Psi_{N}}{ \Psi (\vec{r}_j) } \partial_t \Psi (\vec{r}_j) \\
	&= \frac{1}{N} \sum_{j=1}^{N} \int \diff \vec{r}_j \Psi^* (\vec{r}_j) i \hbar \partial_t \Psi (\vec{r}_j) \\
	& = \int \diff \vec{r} \Psi^* (\vec{r}) i \hbar \partial_t \Psi (\vec{r}) = E_{\Psi}.
\end{split}
\end{equation}
Taking the functional derivative with respect to $ \Psi^* $ yields
\begin{equation}
	i\hbar \partial_t \Psi  = \frac{\delta E_{\Psi}}{\delta \Psi^*},
\label{eq:generalizedGP}
\end{equation}
where the right hand side is given by
\begin{equation}
	\frac{\delta E_{\Psi}}{\delta \Psi^*} = \left(
	-\alpha \nabla^2 + v_\text{ext} +  \int \diff \vec{r}' |\Psi (\vec{r}')|^2 u(\vec{r},\vec{r}') 
	\right) \Psi .
\end{equation}
The functional Gross-Pitaevskii equation  in Eq.~\eqref{eq:generalizedGP} was already discussed by Gross in his seminal paper on superfluid vortices \cite{Gross1961}. This theory corresponds to a quantum density functional theory where the exchange-correlation energy is neglected;  for strongly correlated systems, see e.g. Ref.~\cite{Malet2015}.


\subsection{Fourier expansion for isotropic potentials}

The energy contribution for an isotropic pair potential $ u $ is
\begin{equation}
\begin{split}
\left\langle u \right \rangle &= \frac{1}{2}\int \int \diff{\vec{r}_1} \diff{\vec{r}_2} [ 
\Psi^* (t,\vec{r}_2) \Psi^* (t,\vec{r}_1)
\\ 
&  u(|\vec{r}_2 - \vec{r}_1|) \Psi (t,\vec{r}_1) \Psi (t,\vec{r}_2) 
],
\end{split}
\end{equation}
which can be rewritten as
\begin{equation}
\left\langle u \right \rangle =
\frac{1}{2}
\int \diff \vec{r} \left[
n \varone \left( 
u \ast n
\right) \varone
\right],
\end{equation}
where $ \left( 
u \ast n
\right) $ is the convolution 
\begin{equation}
\int \diff \vec{r}' \left[
u(|\vec{r} - \vec{r}'|) n (t,\vec{r}')
\right]
\end{equation}
and $ n $ is the local number density $ |\Psi|^2 $. 

The Fourier transform of a pair-potential 
\begin{equation}
\hat{u}(k) = \int \diff \vec{r} \left[  
e^{-i \vec{k} \cdot \vec{r}} u (r)
\right] 
\end{equation}
with finite moments can be expressed as a power series in $ k^2 $ as
\begin{equation}
\hat{u}(k) = \sum_{j=0}^{\infty} g_{2j} k^{2j}.
\end{equation}
Using this the convolution of the pair-potential becomes
\begin{equation}
\begin{split}
\left( 
u \ast n
\right)  = \mathcal{F}^{-1} \left[ \hat{u} \hat{n} \right]  
=  \sum_{j=0}^{\infty} (-1)^{j} g_{2j} \nabla^{2j} n .
\end{split}
\end{equation}
Here $ \mathcal{F}^{-1}$ denotes the inverse Fourier transform. 


Now
\begin{equation}
\left\langle u \right \rangle = 	 \frac{1}{2} \int\diff{\vec{r}} \left[
n  \left(
\sum_{j=0}^{\infty} (-1)^{j} g_{2j} \nabla^{2j} 
\right)
n 
\right].
\label{eq:pair-potential_expanded}
\end{equation}
The energy $ E_{\Psi} $ becomes
\begin{equation}
\begin{split}
E_\Psi [\Psi^*, \Psi] &= \langle u \rangle \\ & +
\int \diff \vec{r} \left[
\frac{\hbar^2}{2m} |\nabla \Psi  |^2  + v_\text{ext}  |\Psi |^2 
\right] 
\end{split}
\end{equation}
and the dynamics is given by Eq.~\eqref{eq:generalizedGP}.

The interaction term [Eq.~\eqref{eq:pair-potential_expanded}] can be effectively truncated at low $ j $ because 
the kinetic energy contribution $ \propto |\nabla \Psi|^2 $ penalizes high order Fourier modes forcing their amplitudes to be small. Let us examine the lowest order truncations of  Eq.~\eqref{eq:pair-potential_expanded}:
\begin{itemize}
	\item Zeroth order: This case corresponds to the standard Gross-Pitaevskii equation.
	\item Second order: This case was studied in Ref.~\cite{Veksler2014}. The coefficient $ g_2 $ must be non-negative to ensure that the energy is bounded from below,  thus leading to a penalization of variations in $ n $. As showm in App.~\ref{app:uniform_ground_state},  for repulsive contact interactions ($ g_0 > 0 $) the only possible ground state is the uniform state characterized by a constant density \mbox{$ n \varone = \bar{n} $.}
	\item Fourth order: In App. \ref{sec:ground_states} we show that this order leads to pattern formation.
	\item  Higher orders: Interaction potentials  with multiple competing length scales allow the realization of more complex symmetries, including honeycomb pattern or quasicrystals \cite{Lifshits,Mkhonta2013};  see example in Sec.~\ref{s:quasi-SS}.
\end{itemize}

\subsection{Fourth-order expansion}

Let us reparametrize the contribution due to the interaction potential as
\begin{equation}
\left\langle u \right \rangle 
= \int \diff{\vec{r}} \left[
\frac{u_\text{r}}{2} n ^2 
+ \frac{u_\text{e}}{2} n  \left(
\nabla^2 + q_0^2
\right)^2 n  \right]
.
\label{eq:pair-potential_truncated}
\end{equation}

In terms of the wave function $ \Psi $ we have
\begin{equation}
i \hbar \partial_t \Psi  = \hat{H}
\Psi,
\label{eq:modified_gross-pitaevskii}
\end{equation}
where 
\begin{equation}
\hat{H} = -\frac{\hbar^2}{2 m} \nabla^2 + v_\text{ext} 
+ \left[u_\text{r}  + u_\text{e} \left( \nabla^2 + q_0^2 \right)^2 \right] 
\end{equation}
is the system Hamiltonian.

The parameters are defined as
\begin{equation}
q_0^2 = -\frac{g_2}{2 g_4}, \quad
u_\text{e} = g_4, \quad
u_\text{r} = g_0 - \frac{g_2^2}{4 g_4}.
\end{equation}
For negative values of $ q_0^2 $, no pattern forming is to be expected. The sign of $ q_0^2 $ is fully determined by $ g_2 $ since $ g_4 > 0 $ in order to ensure finite energy for small wavelength Fourier modes.

\section{Units}
In an effort  to make the main text more readable we use some of the same symbols for both SI and reduced units whereas in the Appendices we have tried to maintain mathematical rigor with a possible drawback of having too many symbols in the text. In the main text $ \vec{v} $ denotes the velocity field both in SI units and in units of $ \hbar q_0/m $ whereas in the Appendices the field $ \vec{v} $ is only in units of  $ \hbar q_0/m $. In the Appendices we use the notation $ F_\Psi $ for integral energies, where $ \Psi $ is the field variable. We have omitted the field variable for functionals of the fields $ \rho $ and  $ \vec{v} $. In the main text same upper case characters of the form $ F $ are used for energy integrals both in SI units and reduced units. In both the main text and the Appendices the integral energies that are functionals of reduced fields are also in energy units that are reduced. 
See Tab.~\ref{tab:units} for a partial list of symbols used in the main text and the Appendices. 
\begin{table*}
	\centering
	\caption{Units and notation. Here $ d $ is the dimension of the system.}
	\begin{tabular}{l|l|l|l|l}
		Symbol & Description & Units & Variables & Main text \\
		\hline \hline
		$ \vec{r} $ & length & m &  -  & $ \vec{x} $ \\
		$ t $ & time & s & - & $ t $ \\
		$ \Psi $ & wave function & m$ ^{-d/2} $ & $ \varone $ & $ \Psi $ \\
		$ n $ & number density & m$ ^{-d} $ & $ \varone $ & $ n $ \\
		$ u $ & interaction potential & Jm$ ^d $ & $ r = |\vec{r}| $ & $ u $ \\
		$ \vec{x} $ & length & $ q_0^{-1} $ & - & $ \vec{x} $ \\
		$ \tau $ & time & $ m/(\hbar q_0^2) $ & - & $ t $ \\
		$ s $ & minimization parameter & $ m/(\hbar q_0^2) $ & - & - \\
		$ \Phi $ & wave function & $ \hbar/(q_0 \sqrt{m u_\text{e}}) $ & $ \vartw $ & - \\
		$ \rho $ & number density & $ \hbar^2/(m u_\text{e} q_0^2) $ & $ \vartw $ & $ \rho $ \\
		$ U $ & effective potential energy & $ \hbar^2 q_0^2/m $ & $ \rho $ & $ U $ \\
		$ \vec{u} $ & displacement field & $ q_0^{-1} $ & $ \vartw $ & $ \vec{u} $ \\
	\end{tabular}
	\label{tab:units}
\end{table*}
\section{Hydrodynamic formulation}\label{app:hydrodynamic_formulation}
The wave function $ \Phi $ can be rewritten using the polar decomposition as 
\begin{equation}
\Phi \vartw = R \vartw \exp{\left(i S \vartw \right)}.
\end{equation}
Now $ \rho = |\Phi|^2  = R^2  $. We define $ \vec{v} := \nabla S $ allowing for writing  Eq.~\eqref{eq:modified_gross-pitaevskii} as a set of flow equations
\begin{subequations}
	\begin{equation}
	\label{eq:velocityevolution}
	\dadv{\vec{v}   } = - \nabla \left[Q  + \tilde{v}_\text{ext}  + 
	\left( \alpha_u  + \left( \nabla^2 + 1 \right)^2 \right)  \rho   
	\right] , 
	\end{equation}
	\begin{equation}
	\label{eq:massevolution}
	\partial_\tau \rho  = -\nabla \cdot \left[\rho  \vec{v}  \right],
	\end{equation}
\end{subequations}
where $ \mathrm{D}/\mathrm{D} \tau = \partial_\tau + \vec{v}\cdot \nabla $ is the advective time derivative, $ \alpha_u =  u_\text{r} /(u_\text{e} q_0^4)$ and 
\begin{equation}
Q  = -\frac{1}{2} \frac{\nabla^2 \sqrt{\rho }}{\sqrt{\rho }}
\end{equation}
is the quantum potential. 
Eq.~\eqref{eq:velocityevolution} can be expressed in terms of a functional derivative as
\begin{equation}
\dadv{\vec{v}  } = - \nabla \frac{\delta U}{\delta \rho   },
\end{equation}
where 
\begin{equation}
\begin{split}
U [\rho ] = 	\int   \diff{\vec{x}} & \left[
\frac{|\nabla\rho |^2}{8 \rho} + 
\tilde{v}_\text{ext} \rho +
\frac{\alpha_u}{2} \rho^2 
\right.\\  & \left.
+ \frac{1}{2} \rho \left(
\nabla^2 + 1
\right)^2 \rho \right]
\end{split}
\label{eq:internal_energy}
\end{equation}
is the effective potential energy. 
These dynamics conserve the total energy of the system
\begin{equation}
E[  \rho , \vec{v}  ] = K[\rho, \vec{v}   ] + U[\rho ],
\label{eq:total_energy}
\end{equation}
where 
\begin{equation}
K [\rho , \vec{v}   ] =	\int \diff{\vec{x}} \left[
\frac{1}{2}\rho \vartw |\vec{v}\vartw |^2  \right]
\end{equation}
is the kinetic energy. 

\section{Ground states}\label{sec:ground_states}
In this Section we will discuss ground states of the system described by energy $ E $ [Eq.~\eqref{eq:total_energy}]. We will define dissipative processes that are used to minimize $ E $ and use them to analyze the ground state structure of the system both analytically and numerically.

\subsection{Dissipative dynamics in the density formalism}
The energy $ E $ is \emph{locally} minimized when $ \vec{v}=0 $ and $ U $ is minimized with respect to $ \rho $. Here we present two types of density preserving dissipative flows parametrized by $ s $ that minimize $ U $. 

The first one is given by
\begin{equation}
\begin{split}
&\partial_s \rho \varthr = - \frac{\delta U[\rho]}{\delta \rho \varthr} 
+ \lambda(s)
\\&
= - \left[
Q + \left(
\alpha_u + (\nabla^2 +1) ^2
\right) \rho
\right] + \lambda(s),
\end{split}
\label{eq:non-local_dissipative_dynamics}
\end{equation}
where
\begin{equation}
\lambda = \dashint \diff{\vec{x}} \left[
\frac{\delta U[\rho]}{\delta \rho} 
\right]
\end{equation}
is a Lagrange multiplier that keeps the total density $ \int \rho \diff \vec{x}  $ constant in time. Here $ \dashint $ denotes the integral mean defined by $ \dashint f \diff \vec{x} = \int f \diff \vec{x}/\int \diff \vec{x} $. The process described by Eq.~\eqref{eq:non-local_dissipative_dynamics} will be referred to as \emph{non-local} dissipative dynamics.

\emph{Local} dissipative dynamics is defined by
\begin{equation}
\begin{split}
\partial_s \rho  &= \nabla^2 \frac{\delta U[\rho]}{\delta \rho} 
\\ &
= \nabla^2 \left[
Q + \left(
\alpha_u + (\nabla^2 +1) ^2
\right) \rho 
\right],
\end{split}
\label{eq:localdissipativedynamics}
\end{equation}
It can be shown that both these dynamics will lead to a non-increasing $ U $ i.e. $ \partial_s U(s) \leq 0 $. 

\subsection{Dissipative dynamics in the wave formalism}
Let us define the total  energy in the wave formalism:
\begin{equation}
\label{eq:waveenergy}
\begin{split}
&E_\Phi [\Phi^*,\Phi] =  \\ & \frac{1}{2} \int \diff \vec{x} \left\lbrace
|\nabla \Phi |^2 + \alpha_u |\Phi |^4 
+ \left[
\left(
1 + \nabla^2
\right) 
\left|
\Phi 
\right|^2
\right]^2 
\right\rbrace.
\end{split}
\end{equation}
The generalized nonlinear Schrödinger's equation can be written as
\begin{equation}
\begin{split}
&\partial_s \Phi  = -i \frac{\delta E_\Phi}{\delta \Phi^*} 
\\ &
= -i  \left[-\frac{1}{2} \nabla^2  
+ \left( \alpha_u  +  \left( \nabla^2 + 1 \right)^2 \right) \left| \Phi  \right|^2 
\right]
\Phi  .
\end{split}
\label{eq:waveformalismenergy}
\end{equation}
The time evolution can be made dissipative by making a modification
\begin{equation}
\partial_s \Phi \varthr = - (i + \mu) \frac{\delta E_\Phi}{\delta \Phi^*} + \lambda (s) \Phi \varthr ,
\label{eq:dissipative_wave_formalism_2nd}
\end{equation}
where $ \mu $ is some positive dissipation rate. Here $ \lambda(t) $ is a Lagrange multiplier ensuring the global conservation of the density $ |\Phi|^2 $ and can be calculated as
\begin{equation}
\label{eq:waveformalismlambda}
\lambda (s) = \frac{\mu}{N} \int \diff \vec{x}  \left[
\Phi^* \varthr \frac{\delta E_\Phi}{\delta \Phi^*}  \right]
= \frac{\mu}{N} \langle H \rangle ,
\end{equation}
where
\begin{equation}
N = \int \diff \vec{x} \left[
\left|
\Phi \varthr
\right|^2 \right]
\end{equation}
is the rescaled number of particles in the system and $ H $ is the Hamiltonian operator. This can be interpreted as the expected energy per particle times the dissipation rate. 

The dynamics can be made overdamped by dropping the imaginary part of the mobility giving
\begin{equation}
\partial_s \Phi \varthr = - \mu \frac{\delta E_\Phi}{\delta \Phi^*} + \lambda (s) \Phi \varthr
\label{eq:dissipative_wave_formalism}
\end{equation}
where $ \lambda $ is defined by Eq.~\eqref{eq:waveformalismlambda}. This Equation has no coupling between the real and the imaginary parts of $ \Phi $ allowing for setting $ \Im [\Phi \varthr ]=0 $, which gives $ \Phi \varthr = \sqrt{\rho \varthr}$.

Both Eqs.~\eqref{eq:dissipative_wave_formalism_2nd} and \eqref{eq:dissipative_wave_formalism} can be shown to lead to a non-increasing energy $ E_\Phi $ in time $ s $.

\subsection{Linear stability analysis}\label{app:linear_stability}
Let us analyze the stability of Eq.~\eqref{eq:localdissipativedynamics} against small periodic perturbations about a constant state i.e. $ \rho \vartw = \bar{\rho} + \varepsilon (\tau) \exp{(i \vec{q}\cdot \vec{x})}$, where $ \varepsilon $ is some positive small spatially independent amplitude. Inserting this ansatz in Eq.~\eqref{eq:localdissipativedynamics} and abbreviating $ q = |\vec{q}| $ we find
\begin{equation}
\begin{split}
\partial_\tau \varepsilon (\tau) &= -\left(\frac{q^4}{4 \bar{\rho}} + \alpha_u q^2 + q^2 (1-q^2)^2\right)\varepsilon(t) \\
&= C(q) \varepsilon(\tau)
\end{split}
\end{equation}
up to linear order in $ \varepsilon $.
Maximizing the coefficient $ C $ with respect to $ q $ define the most \emph{unstable} perturbation. If such maximum exists, it suffices to examine the value of $ C $ at the given maximum. If no maximum exists, $ C $ is maximized by $ q=0 $, which will give a stable constant solution. 

\begin{figure}
	\centering
	\includegraphics[width = 3in]{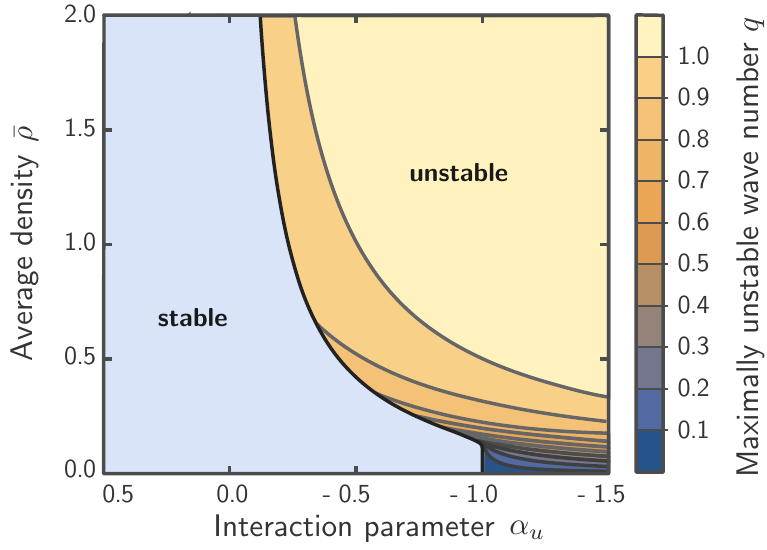}
	\caption{Stability diagram of the uniform density state as a function of the interaction parameter $ \alpha_u $ and average density $ \bar{\rho} $. The contour lines show the value for the most unstable wave number $ q $ of the linear perturbation. }
	\label{fig:stability_analysis}
\end{figure}

Solving the aforementioned problem gives the condition for stability. The uniform solution is \emph{not} stable against perturbations if
\begin{equation}
\alpha_u <
\begin{cases}
-1, & \bar{\rho} \leq 1/8,  \\ 
(1-16 \bar{\rho})/(64 \bar{\rho}^2),& \bar{\rho}>1/8.
\end{cases}
\end{equation}
%
This defines the spinodal curve for the uniform phase. 
The stability phase diagram with the value for the most unstable $ q $ is shown in Fig~\ref{fig:stability_analysis}. 

\subsection{One-mode approximation} \label{app:one-mode_approximation}
The pattern forming part of the internal energy Eq.~\eqref{eq:internal_energy} can be written in Fourier space as 
\begin{equation}
\frac{1}{8 \pi^2} \int \diff \vec{k} 
\left[
(1-k^2)^2 \hat{\rho}(\vec{k})^2 
\right].
\end{equation}
by using Plancherel's theorem. 
For crystalline ground states we have
\begin{equation}
\rho (\vec{x}) = \sum_{\vec{k} \in G} a_\vec{k} e^{i \vec{k} \cdot \vec{x}},
\end{equation}
where $ G $ is the point group of the reciprocal crystal lattice. 
The pattern forming  term will penalize any modes with $ |\vec{k}| \neq 1 $
suggesting  that including  only the vectors  $ \vec{k} $ in the first Brillouin zone might approximate well the ground state. 
Here we analyze 
one-mode approximations of periodic number densities in two dimensions.

\subsubsection{Stripe phase}

We use the ansatz
\begin{equation}
\begin{split}
\rho_{\text{S}}(\vec{x}) &= \bar{\rho} +\phi_0 n_{\text{S}}(\vec{x})= \bar{\rho} + \phi_0 \left(
e^{i \vec{q} \cdot \vec{x}} + \text{c.c.}  
\right) \\
&= \bar{\rho}\, (1 + A \cos{(\vec{q} \cdot \vec{x})} ),
\end{split}
\label{eq:stripe_ansatz}
\end{equation}
where $ \text{c.c.} $ stands for the \emph{complex conjugate} and $ A = 2 \phi_0/\bar{\rho} $. We can simplify the calculations by choosing $ \vec{q} = (q,0) $ and writing $ \vec{q} \cdot \vec{x} = qx $ leading to 
\begin{equation}
\begin{split}
\notag
\rho_{\text{S}}(x) &= \bar{\rho} +\phi_0 n_{\text{S}}(x)= \bar{\rho} + \phi_0 \left(
e^{i qx} + \text{c.c.}  
\right) \\
&= \bar{\rho}\, (1 + A \cos{(qx)} ).
\end{split}
\label{eq:stripe_ansatz_1d}
\end{equation}
Inserting the stripe ansatz $ \rho_{\text{S}} $ in Eq.~\eqref{eq:internal_energy} gives an average energy
\begin{equation}
\begin{split}
&\bar{U} = \\ &\dashint_{C_p} \diff{x} \left[
\frac{1}{8 }  \frac{|\nabla \rho_{\text{S}}  |^2}{\rho_{\text{S}} } + 
\frac{\alpha_u}{2} \rho_{\text{S}}  ^2 
+ \frac{1}{2} \rho_{\text{S}} \left(
\nabla^2 + 1
\right)^2 \rho_{\text{S}}  \right],
\end{split}
\end{equation}
where $ C_p $ is the interval $ [0,2\pi/q] $ corresponding to a period of $ \rho_{\text{S}} $. The integrand is constant in the $ y $-coordinate reducing the energy to a one-dimensional integral. We start by calculating the quadratic term
\begin{equation}
\notag
\rho_{\text{S}}^2 = \bar{\rho}^2 + 2 \bar{\rho} \phi_0 n_{\text{S}}(x) +\phi_0^2 n_{\text{S}}(x)^2.
\end{equation}
All the terms that have an oscillating component $ \exp(i q j x ) $ with $ j=1,2,3... $ give no contribution to the average energy. Therefore the interesting terms are $ n_\text{S}^p $ with $ p>1 $. The terms without an oscillating component will be referred to as \emph{resonant} terms since they resonate with the linear operation $ \int_{C_p} \diff \vec{x} $.

Now
\begin{equation}
\notag
\dashint_{C_p} \diff{x} \left(
n_{\text{S}}^2 \right) =  \dashint_{C_p}  \diff{x}  \left(      
e^{2iqx} + e^{-2iqx} + 2
\right)  = 2
\end{equation}
and 
\begin{equation}
\notag
\dashint_{C_p} \diff x \left[
\rho_{\text{S}}(x)^2 \right] = \bar{\rho}^2 + 2\phi_0^2.
\end{equation}
We also need
\begin{equation}
\begin{split}
\notag
\dashint_{C_p} &\diff x \left[
\rho_{\text{S}}(x) \nabla^{2p} \rho_{\text{S}}(x) \right] 
\\ &= 
\dashint_{C_p} \diff x \left[
(-1)^p \rho_{\text{S}}(x) q^{2p} n(x)
\right] \\
&= 2 (-1)^p q^{2p} \phi_0^2.
\end{split}
\end{equation}
The pattern forming part of the average energy becomes 
\begin{equation}
\begin{split}
\dashint_{C_p} &\diff x \left[
\frac{\alpha_u}{2} \rho_{\text{S}} (x) ^2 
+ \frac{1}{2} \rho_{\text{S}}(x) \left(
\nabla^2 + 1
\right)^2 \rho_{\text{S}} (x) \right]
\\
&= 
\frac{\alpha_u + 1}{2} \bar{\rho}^2 + \left(  
(1-q^2)^2 + \alpha_u
\right)\phi_0^2.
\end{split}
\label{eq:pattern_forming_part}
\end{equation}
The remaining part of the average energy is the part from the quantum potential
\begin{equation}
\notag
\dashint_{C_p} \diff x \left[
\frac{|\nabla \rho_{\text{S}} (x) |^2}{\rho_{\text{S}} (x)} \right]
=\dashint_{C_p} \diff x \left[
\frac{(\partial_x n_\text{S} (x) )^2}{\bar{\rho} + n_\text{S}(x)}
\right]
\end{equation}
Here we will use $ n_\text{S}(x) = \bar{\rho} A \cos{\xi} $, where $ \xi = q x $. 
\begin{equation}
\begin{split}
\notag
&\dashint_{C_p} \diff x \left[
\frac{(\partial_x n_\text{S} (x) )^2}{\bar{\rho} + n_\text{S}(x)}
\right] \\
&=\frac{ A^2 q^2 \bar{\rho}}{2 \pi} \int_0^{2\pi} \diff \xi \left[
\frac{\sin{^2\xi}}{1 + A \cos{\xi}} \right]
\\
&= \frac{ A^2 q^2 \bar{\rho}}{2 \pi} \int_{-\pi}^{\pi}  \diff \xi \left[
\frac{\sin{^2\xi}}{1 + A \cos{\xi}} \right].
\end{split}
\end{equation}
\begin{figure}
	\centering
	\centering
	\includegraphics[width=2.27in]{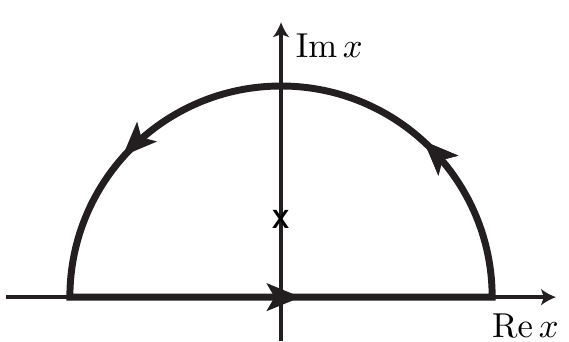}
	\caption{}{
		Semicircular contour in the complex plane used for calculating an integral on the real axis. The symbol $ \bm{\mathsf{x}} $ marks a singularity on the imaginary axis.
	}
	\label{fig:contour}
\end{figure}
\begin{figure}
	\centering
	\includegraphics[width=1.97in]{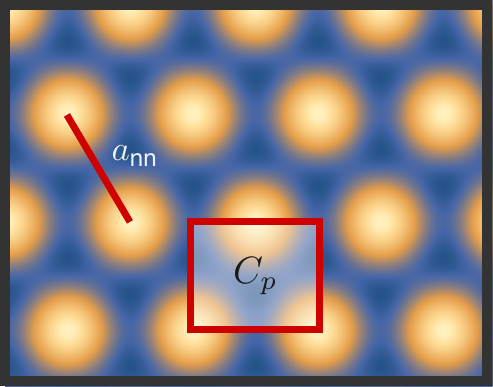}
	\caption{}{
		One-mode density field with a hexagonal symmetry showing the primitive lattice cell $ C_p $ and the nearest neighbor distance $ a_\text{nn} = 4\pi/ (q \sqrt{3}) $.
	}
	\label{fig:hexagonal_field}
\end{figure}
This can be solved with a substitution $ p = \tan{(\xi/2)} $. This gives
\begin{equation}
\notag
\sin{\xi} = \frac{2 p}{1+p^2}, \quad
\cos{\xi} = \frac{1-p^2}{1+p^2}, \quad
\diff \xi = \frac{2}{1+p^2} \diff p.
\end{equation}
Now,
%
\begin{equation}
\notag
\begin{split}
&\frac{ A^2 q^2 \bar{\rho}}{2 \pi} \int_{-\pi}^{\pi} \diff \xi \left[
\frac{\sin{^2\xi}}{1 + A \cos{\xi}}  \right] 
\\
&= 
\frac{ A^2 q^2 \bar{\rho}}{2\pi} \int_{-\infty}^{\infty} \diff p \left[
\frac{\left( \frac{2p}{1+p^2}\right)^2}{1 + A\left( \frac{1-p^2}{1+p^2} \right)} \frac{2}{1+p^2} 
\right] \\
&= \frac{8  A^2 q^2 \bar{\rho}}{2 \pi (1-A)} \int_{-\infty}^{\infty} \diff p \left[
\frac{p^2}{(p^2 + \gamma_A^2)\left( 1+p^2\right)^2} 
\right]
,
\end{split}
\end{equation}
%
where $ \gamma_A^2 = (1+A)/(1-A) $. Note that $ \rho_{\text{S}} \geq 0 $ requires $ A\leq 1 $. We also assume $ A \geq 0 $ implying that  $ \gamma_A^2 > 0  $ and that $ \gamma_A $ is real.  The integral
\begin{equation}
\notag
\begin{split}
& \int_{-\infty}^{\infty} \diff p \left[
\frac{p^2}{p^2 + \gamma_A^2} \frac{1}{\left( 1+p^2\right)^2} 
\right] \\ &
=  \int_{-\infty}^{\infty} \diff p \left[
\frac{p^2}{(p+i \gamma_A)(p-i \gamma_A)} \frac{1}{(p+i)^2 (p-i)^2 } \right] \\&
=: \int_{-\infty}^{\infty} \diff p \,f(p) 
\end{split}
\end{equation}
can be solved in a standard way using residue theorem for a semicircular path on the complex plane shown in Fig.~\ref{fig:contour} and taking the radius to $ \infty $. For this we need residues at singularities in the upper-half complex plane ($ \Im{p} > 0 $). There are two of these, $ p=i \gamma_A $ and $ p = i $. The latter is of order two. 

\begin{equation}
\notag
\begin{split}
&\operatorname{Res}{(f,i \gamma_A)} = \frac{i \gamma_A }{2 (\gamma_A^2-1)^2},  \\
&\operatorname{Res}{(f,i)} = - \frac{i (1+  \gamma_A^2) }{4 (\gamma_A^2-1)^2}.  \\ &
\vphantom{k}
\end{split}
\end{equation}
Now,
\begin{equation}
\notag
\int_{-\infty}^{\infty} 	 \diff p \,
f(p) 
= 2\pi i \left[
\operatorname{Res}{(f,i \gamma_A)} + \operatorname{Res}{(f,i)} 
\right]
\end{equation}
giving 
\begin{equation}
\notag
\begin{split}
&\frac{8  A^2 q^2 \bar{\rho}}{2 \pi (1-A)} \int_{-\infty}^{\infty} \diff p \left[
\frac{p^2}{p^2 + \gamma_A^2} \frac{1}{\left( 1+p^2\right)^2} 
\right] \\
&= q^2 \bar{\rho} \left(
1 - \sqrt{1-A^2}
\right).
\end{split}
\end{equation}

Finally,
\begin{equation}
\begin{split}
\bar{U} &= \frac{q^2 \bar{\rho}}{8} \left(
1 - \sqrt{1-4 \left(\frac{\phi_0}{\bar{\rho}} \right)^2 } \right) \\&
+ \frac{\alpha_u + 1}{2} \bar{\rho}^2 + \left(  
(1-q^2)^2 + \alpha_u
\right)\phi_0^2. \\&
\vphantom{k^2}
\end{split}
\label{eq:onemode_stripe_energy}
\end{equation}

\subsubsection{Hexagonal phase}
\label{sec:hexagonal_phase}
The crystalline hexagonal ground-state solution is approximated similarly to the stripe phase with the one-mode approximation
\begin{equation}
\begin{split}
\rho_\text{H}(\vec{r}) &= \bar{\rho} + \phi_0 n_{\text{H}}(\vec{r}) \\ &
= \bar{\rho} + \phi_0 \sum_{j=1}^{3}\left(
e^{i \vec{q}_j \cdot \vec{r}} + \text{c.c.} 
\right),
\end{split}
\label{eq:hexagonal_ansatz}
\end{equation}
where the reciprocal lattice vectors $ \vec{q}_j $ have a hexagonal symmetry i.e. 
\[
\vec{q}_j \cdot \vec{q}_i = 
\begin{cases}
q^2 & i=j \\
-\frac{1}{2} q^2 & i\neq j.
\end{cases}
\]
%

$ \vphantom{k^2} $

Equation \eqref{eq:hexagonal_ansatz} can be rewritten as 
\begin{equation}
\notag
\begin{split}
&\rho_\text{H} = \\
&\bar{\rho} \left(
1 + 2 A \cos{\left( \frac{\sqrt{3} qx}{2} \right)} \cos{\left(\frac{qy}{2} \right)} + A \cos{(qy)}
\right),
\end{split}
\end{equation}
where, again, $ A = 2 \phi_0/\bar{\rho} $.

Calculating the resonant terms for the polynomial part of the internal energy is pretty straightforward \cite{Provatas2010}. For the quantum potential we have

\begin{widetext}
	\begin{equation}
	\notag
	\dashint_{C_p} \diff \vec{x} \left[
	\frac{1}{8} \frac{|\nabla \rho_\text{H}|^2}{\rho_\text{H}} \diff \right] = 
	\dashint_{C_p} \diff \vec{x} \left[
	\frac{\bar{\rho} q^2 A^2}{8} \frac{3  \sin ^2\left(\frac{\sqrt{3} qx}{2}\right) \cos ^2\left(\frac{qy}{2}\right)+ \left( \cos \left(\frac{\sqrt{3} qx}{2}\right) \sin \left(\frac{qy}{2}\right)+  \sin (qy)\right)^2}{2 A \cos \left(\frac{\sqrt{3} qx}{2}\right) \cos \left(\frac{qy}{2}\right)+A \cos (qy)+1} \right] ,
	\label{eq:hexqpstart}
	\end{equation}
	where $ C_p $ is the primitive lattice cell shown in Fig.~\ref{fig:hexagonal_field}.
	Let $ \sqrt{3}qx/2 = \tilde{x} $ and $ qy = \tilde{y} $. Now this integral is
	\begin{equation}
	\notag
	\frac{\bar{\rho} q^2 A^2}{32 \pi^2} \int_{0}^{2\pi} \int_{0}^{2\pi} \diff \tilde{x} \diff \tilde{y} \left[
	\frac{3  \sin ^2\left(\tilde{x} \right) \cos ^2\left(\frac{\tilde{y}}{2}\right)+\left( \cos \left(\tilde{x} \right) \sin \left(\frac{\tilde{y}}{2}\right)+ \sin (\tilde{y})\right)^2}{2 A \cos \left( \tilde{x} \right) \cos \left(\frac{\tilde{y}}{2}\right)+A \cos (\tilde{y})+1} \right]. 
	\end{equation}

	The integration in $ \tilde{x} $ can be done by repeating the calculation done for the stripe phase. The integrand will have two purely imaginary singularities that can be calculated by factoring the polynomial appearing in the denominator. This gives an integral in $ \tilde{y} $ over a period of $ 2\pi $
	\begin{equation}
	\notag
	\begin{split}
	\frac{\bar{\rho} q^2}{32 \pi}\int_{-\pi}^{\pi} \diff \tilde{y} & \left[
	\frac{1}{\cos (\tilde{y})+1} \left(
	\frac{-2 -6 A + 10 A^2 +\left(9 A^2-4 A-4\right) \cos (\tilde{y})-2 A \cos (2 \tilde{y}) - A^2 \cos (3 \tilde{y}) }{ \sqrt{2 A^2 \cos (2 \tilde{y})-6 A^2-8 (A-1) A \cos (\tilde{y})+4}} \right. \right.  \\ & \qquad \left. \left.
	+ 
	(A+2) \cos (\tilde{y})+2 A+1
	\vphantom{\frac{A^2 (-\cos (3 \tilde{y}))+\left(9 A^2-4 A-4\right) \cos (\tilde{y})+10 A^2-2 A \cos (2 \tilde{y})-6 A-2}{ \sqrt{2 A^2 \cos (2 \tilde{y})-6 A^2-8 (A-1) A \cos (\tilde{y})+4}}}
	\right) \right]
	.
	\end{split}
	\end{equation}
\end{widetext}
Noticing that $ \cos (-\tilde{y}) = \cos (\tilde{y}) $ and making the change of variables $ \cos{\tilde{y}} = 1 - 2 \xi $ gives
\begin{equation}
\notag
\frac{q^2\bar{\rho}}{32 \pi } \int_0^1 \diff \vec{\xi}
\left[ 
\frac{P_1(\xi)}{\xi \sqrt{f_1}} 
+ \frac{Q_1(\xi)}{\xi \sqrt{\xi (1-\xi)}}
\right]
,
\end{equation}
where 
\begin{equation}
\notag\begin{split}
P_1 (\xi) 
&= -16 A^2 \xi ^3+\left(24 A^2-8 A\right) \xi ^2 \\ & +(4 A-4) \xi + (A-1)^2,
\end{split}
\end{equation}
\begin{equation}
\notag
Q_1 (\xi) = (2 A+4) \xi +A-1,
\end{equation}
and
\begin{equation}
\notag\begin{split}
&f_1(\xi) =\\& (4 A^2 \xi ^2+\left(4 A-8 A^2\right) \xi +A^2-2 A+1)(1-\xi)\xi .
\end{split}
\end{equation}

Let us first calculate the part
\begin{equation}
\notag\begin{split}
&I_1 := \int_{0}^{1} \diff \xi \left[
\frac{Q_1(\xi)}{\xi \sqrt{\xi (1-\xi)}}
\right] \\&
= \int_{0}^{1} \diff \xi \left[
\frac{4 + 4 A}{ \sqrt{\xi (1-\xi)}}
\right] - 
\int_{0}^{1} \diff \xi \left[
\frac{1 - A}{ \xi \sqrt{\xi (1-\xi)}}
\right]
\end{split}
\end{equation}
Both parts can be calculated with trigonometric substitution. This gives
\begin{equation}
\begin{split}
I_1 &= 2(2 + A) \pi - \lim\limits_{\xi \to 1} \frac{2(1 - A)\xi}{\sqrt{\xi(1-\xi)} } \\ &
= 2(2 + A) \pi - \lim\limits_{\epsilon \to 0} 2(1 - A)\sqrt{\epsilon^{-1}-1} .
\end{split}
\end{equation}
The second part diverges and has to be combined with other terms. 
Now the other part needed here is
\begin{equation}
\begin{split}
&I_2 := \int_0^1 \diff \xi \left[
\frac{P_1(\xi)}{\xi \sqrt{f_1}}
\right] \\ &
= \int_0^1 \diff \xi \left[
-\frac{16 A^2 \xi ^2}{\sqrt{f_1}} - \frac{8A \left( 1 - 3  A \right) \xi}{\sqrt{f_1}} 
\right. \\ & \left.
-  \frac{4 (1-A)}{\sqrt{f_1}}  + \frac{(1-A)^2}{\xi \sqrt{f_1}}
\right].
\end{split}
\end{equation}
These are elliptic integrals. The only diverging part is 
\begin{equation}
\notag
\int_{0}^{1} \diff \xi \left[
\frac{(1 - A)^2}{\xi \sqrt{f_1}} 
\right]
.
\end{equation}
This can be dealt with by using an identity 
\begin{equation}
\begin{split}
&	(2-s)a_0 J_{s-3} + \frac{1}{2} a_1 (3 - 2s) J_{s-2} \\ &
+ a_2 (1-s) J_{s-1} 
+ \frac{1}{2} a_3 (1 - 2s ) J_s - s a_4 J_{s+1} \\ &
= \left. \sqrt{f} (\xi - c)^{-s}\right\rvert_0^1 ,
\end{split}
\end{equation}
where 
\begin{equation}
J_s[f] = \int_0^1 \diff \xi \left[
\frac{1}{\sqrt{f} (\xi - c )^s}
\right],
\end{equation}
$ f $ is a third or fourth order polynomial with
\begin{equation}
\notag
\begin{split}
f &= a_0 (x - c )^4 + a_1 (x - c)^3 \\ &
+ a_2 (x - c)^2 + a_3 (x - c) + a_4,
\end{split}
\end{equation}
$ s = 1, 2,3,... $ and $ c $ is some arbitrary constant \cite{abramowitz+stegun}. In our case $ s = 1 $, $ c=0 $ and  $ f = f_1 $
giving 
\begin{equation}
\notag
\begin{split}
&a_0 = - 4 A^2, \;
a_1 = - 4 (1 - 3 A) A, \; 
a_2 = -(1 - 3 A)^2, \\&
a_3 = (1 - A)^2, \;
a_4 = 0.
\end{split}
\end{equation}
We have
\begin{equation}
\notag
\begin{split}
&\int_{0}^{1} \diff \xi \left[
\frac{(1 - A)^2}{\xi \sqrt{f_1}} 
\right] =
-(1 - A)^2 J_1 \\ &
= -8 A^2  J_{-2} 
+ 4 A (3 A-1)  J_{-1} 
+ 
\left. \frac{-2 \sqrt{f_1}}{ \xi} \right\rvert_0^1  .
\end{split}
\end{equation}
The only diverging part is the last one. Combining with the earlier diverging term gives
\begin{equation}
\notag
\begin{split}
&\lim\limits_{\epsilon \to 0} \left(	\left. \frac{-2 \sqrt{f_1}}{\xi} \right\rvert_\epsilon^1  - 2(1 - A)\sqrt{\epsilon^{-1}-1 }   \right) \\&
= -2 \sqrt{f_1(1)} = 0.
\end{split}
\end{equation}
Let 
\begin{equation}
I_t = I_1 + I_2.
\end{equation}
%
Now
\begin{equation}
\notag
\begin{split}
I_t &= 2(2+A)\pi 
-4 (1 - A) J_0 \\&
- 12 A (1 - 3 A) J_{-1}
- 24 A^2 J_{-2}. 
\end{split}
\end{equation}
The integrals $ J_s $ are standard elliptic integrals and can be transformed to Legendre normal form with a change of integration variables. We have $ f_1 = g_1 g_2 $, where $ g_2 = \xi (1-\xi) $ and $ g_1(\xi) > 0 $ for $ \xi \in [0,1] $. We use the change of variables $ \xi_2 = \sqrt{g_2/g_1} $.  For details, see Ref.~\cite{abramowitz+stegun}. The contribution of the quantum potential can be written as $ q^2 \bar{\rho} I_t /32 \pi $,
which, after a somewhat tedious calculation, gives
\begin{widetext}
	\begin{equation}
	U_{\text{QP}} =
	\frac{q^2 \bar{\rho}}{16}
	\left\lbrace
	A+2
	-
	\sqrt{2-3 A}
	\left[
	\frac{
		12 C_A^{-1} \left(E\left(k_A\right)-K\left(k_A\right)\right)+(2 A+1) C_A K\left(k_A\right)
	}{
		\pi}
	\right]
	\right\rbrace ,
	\label{eq:hexfinalsolution}
	\end{equation}
\end{widetext}
%
where 
\begin{equation}
K(k) = \int_{0}^{1} \diff t \left[
\frac{1}{\sqrt{(1-t^2)(1-k^2 t^2) }} \right]
\end{equation}
is the complete elliptic integral of the first kind
and
\begin{equation}
E(k) = \int_{0}^{1} \diff t \left[
\frac{\sqrt{1-k^2 t^2}}{\sqrt{1-t^2}} \right]
\end{equation}
is the complete elliptic integral of the second kind. The complex coefficients are defined as
\begin{equation}
k_A = i \sqrt{\frac{3 A^2+\sqrt{(1-A)^3 (3 A+1)}-1}{-3 A^2+\sqrt{(1-A)^3 (3 A+1)}+1}}
\end{equation}
%
and
\begin{equation}
C_A = \sqrt{\frac{6 A^2+2 \sqrt{(1-A)^3 (3 A+1)}-2}{A^3}}.
\end{equation}
The value of $ U_{\text{QP}} $ at the maximal $ A = 2/3 $ can be expressed in terms of known constants as
\begin{equation}
\left. U_{\text{QP}} \right\rvert_{A=2/3} = \left(\frac{1}{6}-\frac{\sqrt{3}}{8 \pi } \right) q^2 \bar{\rho}.
\end{equation}
For more details see the Mathematica notebook \textsf{elliptic\_integral.nb}.

\begin{figure}
	\centering
	\includegraphics[width=3in]{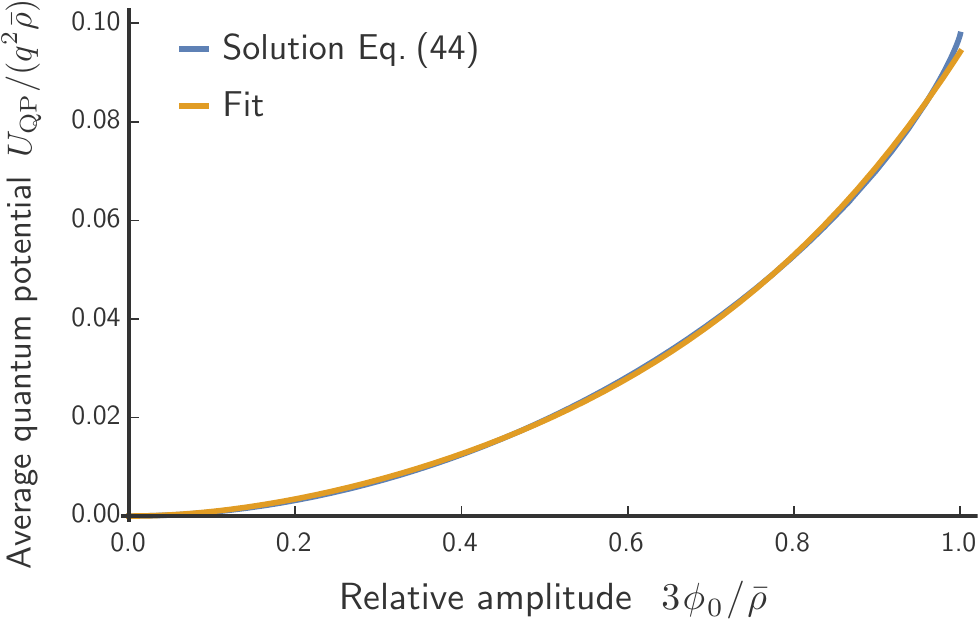}
	\caption{
		The exact solution for the one-mode approximation energy of the quantum potential $ U_\text{QP} $ plotted against a polynomial fit showing that the quartic approximation suffices for analytical calculations. 
	}
	\label{fig:quantum_potential_approximation}
\end{figure}

$ U_{\text{QP}} $ will be a part of the total internal energy that still has to be minimized with respect to $ q $ and $ A $. To simplify the minimization procedure we fit a fourth order polynomial
\begin{equation}
p_4(x) = a_2 A^2 + a_3 A^3 + a_4 A^4
\end{equation}
against  the quantum potential \eqref{eq:hexfinalsolution}. We find
\begin{equation}
\begin{split}
\bar{U} &= (a_2  + a_3 A + a_4 A^2) A^2 q^2 \bar{\rho} \\&
+ \frac{1}{4} \left(
2 (\alpha_u +1)+3 A^2 \left(\alpha_u +(q^2-1)^2 \right) \right)
\bar{\rho}^2,
\end{split}
\label{eq:hexagonal_onemode_energy}
\end{equation}
where $ a_2 = 0.226034 $, $ a_3 = -0.288956 $, and $ a_4 = 0.401767 $.  The zeroth coefficients disappears because for zero amplitude the quantum potential is zero. On the other hand, the first coefficient is zero because the integral over the periodic domain of any single oscillating mode gives zero. Fourth order fit seems to agree well with the exact solution as shown in Fig.~\ref{fig:quantum_potential_approximation}.

\subsection{Finding the ground state}
\begin{figure}
	\centering
	\includegraphics[width=3.in]{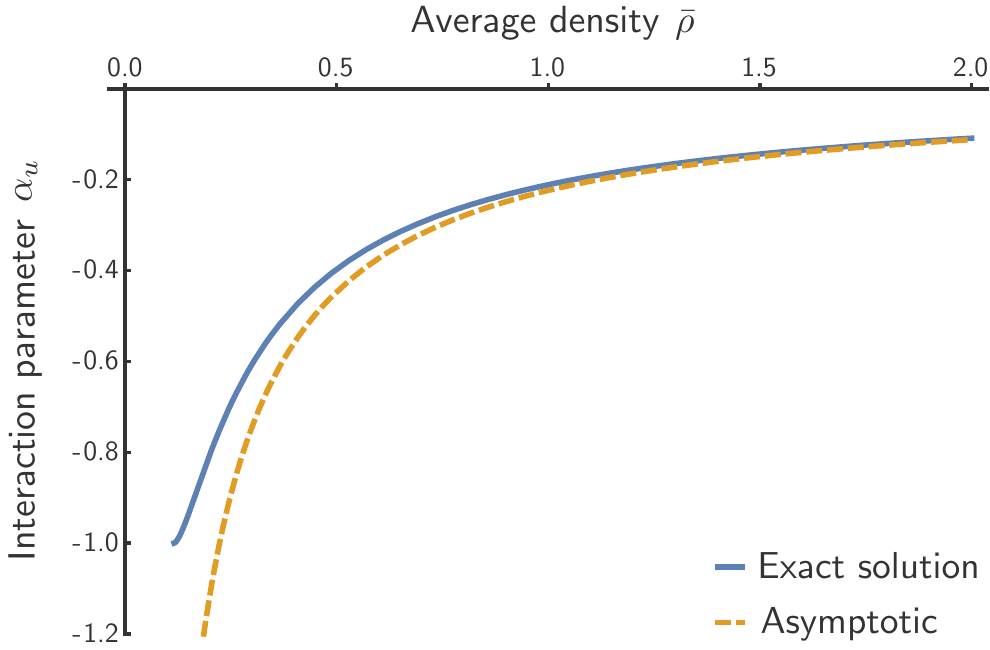}
	\caption{
		The the uniform--supersolid phase transition line with an asymptotic expression $ \alpha_u = -0.2234/\bar{\rho} $.
	}
	\label{fig:phase_transition_line_asymptotic}
\end{figure}

The ground state of the system is found by minimizing Eqs. \eqref{eq:onemode_stripe_energy} and  
\eqref{eq:hexagonal_onemode_energy} with respect to the scaler $ q $ and the amplitude $ A = 2\phi_0/\bar{\rho} $ for given system parameters $ \bar{\rho} $ and $ \alpha_u $. In both of the cases, the extrema can be expressed as roots of polynomials. The coexistence gap between the uniform superfluid and the supersolid phases is calculated numerically using the analytical energies via common tangent construction. We find an analytical expression for the phase transition line between the uniform superfluid and the supersolid phases that can be expressed as an asymptotic expansion for large $ \bar{\rho} $ as $ \alpha_u = -0.2234/\bar{\rho}  $. Fig.~\ref{fig:phase_transition_line_asymptotic} shows the asymptotic expression with the exact solution.
For details on the ground state phase diagram see the Mathematica notebook \textsf{phase\_diagram.nb}.

\section{Wave equation for lattice vibrations of the supersolid }\label{app:wave_equation}
We next examine the dynamics of small displacement field about the hexagonal ground state described in Sec.~\ref{sec:hexagonal_phase}. To this end, we will introduce a deformation of the coordinates $ \vec{x} \to \vec{x} - \vec{u} \vartw $. The one-mode approximation in Eq.~\eqref{eq:hexagonal_ansatz} becomes
%
%
\begin{equation}
\rho\vartw = \bar{\rho} + \phi_0 \sum_{j=1}^3 \left[
e ^{-i\vec{q}_j \cdot \vec{u}\vartw }  e^{i \vec{q}_j \cdot \vec{x}} + \text{c.c.}
\right].
\label{eq:density_amplitude}
\end{equation}
We will write down the time evolution equation for the displacement field $ \vec{u} \vartw $ in terms of the effective potential energy $ U $. First, we assume that the relative amplitude $ \phi_0/\bar{\rho} $ is small and expand the quantum potential up to fourth order in $ \phi_0/\bar{\rho} $. This assumption should work near the uniform--supersolid phase transition line, where the amplitude of the oscillating number density is small. Second, we assume that $ \nabla \vec{u} $ is small and expand the configuration energy $ U $ up to second order in $ \vec{u} $. Third, we assume that the velocity field $ \vec{v} $ is small and keep only the  linear terms in the dynamical equations. 
For details of the calculation see Ref.~\cite{Heinonen2016_2}, where a similar analysis was performed. 

Now, Eqs.\eqref{eq:velocityevolution} and \eqref{eq:massevolution} give
\begin{equation}
\label{eq:displacementdynamics}
\partial_\tau \vec{u}\vartw = \vec{v}\vartw, \quad
\partial_\tau \vec{v}\vartw = -  \frac{1}{\bar{\rho}} \frac{\delta U[\vec{u}] }{\delta \vec{u}},
\end{equation}
assuming that $ \phi_0 $ and $ |\vec{q}_j| = q $ take the equilibrium value. This can be combined into a single equation
\begin{equation}
\partial_\tau^2 \vec{u}\vartw = -   \frac{1}{\bar{\rho}} \frac{\delta U[\vec{u}] }{\delta \vec{u}}.
\label{eq:generalwaveeq}
\end{equation}
The configuration energy $ U $ with the aforementioned approximations reduces to
\begin{equation}
\begin{split}
&U [\vec{u}\vartw] =
\frac{3\phi_0^2 q^2 }{2} \int \diff \vec{x}  \left\lbrace \vphantom{\frac{\phi_0^2}{\bar{\rho}^2}}
\right. \\ &\left.
\left[
\frac{1 }{4 \bar{\rho}} 
\left(
1 - \frac{\phi_0}{\bar{\rho}} + 5 \frac{\phi_0^2}{\bar{\rho}^2}
\right) + (3 q^2 - 2 ) 
\right]
\norm{\nabla \vec{u}}^2
\right. \\ &\left.
+ q^2 \left(
(\nabla \cdot \vec{u})^2 + (\nabla \vec{u})^T : \nabla \vec{u} 
\right)
\vphantom{\frac{\phi_0^2}{\bar{\rho}^2}} 
+\frac{2}{3}  \norm{\nabla^2 \vec{u}}^2
\right\rbrace 
\\ &
+ U_0(\alpha_u,\phi_0,\bar{\rho},q).
\end{split}
\label{eq:smalldisplacementenergy}
\end{equation}
Taking the functional derivatives with respect to $ \vec{u} $ and plugging in
Eq.~\eqref{eq:generalwaveeq} leads to
\begin{equation}
\begin{split}
&\partial_\tau^2 \vec{u} =  \frac{3 \phi_0^2 q^2}{\bar{\rho}} \left[
2 q^2 \nabla (\nabla \cdot \vec{u}) 
\right. \\ &\left.
+ \left(
3 q^2 - 2 + \frac{1-\phi_0/\bar{\rho} + 5 \phi_0^2 / \bar{\rho}^2}{4 \bar{\rho}}
\right) \nabla^2 \vec{u} 
-\frac{2}{3} \nabla^4 \vec{u}
\right].
\end{split}
\end{equation}
Inserting a plane wave $ \exp{[i( \vec{k} \cdot \vec{x}-\omega \tau)]} $ gives a dispersion relation 
\begin{equation}
\omega/\omega_\parallel = k/k_\parallel \sqrt{1+(k/k_\parallel)^2}
\end{equation}
for 
longitudinal modes. Here 
%
%
%
\begin{equation}
\begin{split}
k_\parallel^2 &= \frac{3}{2} \left[(5 \phi_0^2/\bar{\rho} ^2-\phi_0/\bar{\rho} +1)/4 \bar{\rho} - 2 + 5 q^2  \right], \\
\omega_\parallel^2 &=  2 q^2 \phi_0 ^2 k_\parallel^4/\bar{\rho}.
\end{split}
\end{equation}

At small $ k $ the dispersion relation becomes linear and one can read off the speed of sound:
\begin{equation}
c_\parallel ^2 = 
\frac{ 3 \phi_0^2 q^2}{\bar{\rho}}
\left(
5 q^2 - 2 + \frac{1-\phi_0/\bar{\rho} + 5 \phi_0^2 / \bar{\rho}^2}{4 \bar{\rho}}
\right).
\end{equation}

We can also calculate $ U_0 $ of Eq.~\eqref{eq:smalldisplacementenergy} giving
\begin{equation}
U_0 = 3 \phi_0^2 V_\Omega \left[
\frac{q^2}{4\bar{\rho}} (1-\phi_0/\bar{\rho} + 5 \phi_0^2/\bar{\rho}^2) + (1-q^2)^2
\right],
\end{equation}
where $ V_\Omega $ is the volume of the domain. Since $ \phi_0/\bar{\rho} \leq 1/3 $, the first part due to quantum potential penalizes any system with $ q>0 $ making the system expand. The second part penalizes any variance from $ q=1 $. This part stabilizes the non-uniform pattern. For large values of $ \bar{\rho} $ the pattern forming part dominates and $ q \approx 1 $. 

%
%





\section{No pattern forming for low level expansions with repulsive contact interaction}\label{app:uniform_ground_state}

Here we will show that the low level expansions of the interaction term $ u $ will only give uniform ground state solutions. The effective potential energy functional for the 2nd order expansion is 
\begin{equation}
U_n[n] = \int  \diff \vec{r} \left[
\frac{\hbar^2}{8m} \frac{|\nabla n |^2}{n } + \frac{g_0}{2} n  ^2 + \frac{g_2}{2} |\nabla n |^2 \right],
\end{equation}
where we use the subscripted $ U_n $ here to emphasize that the energy is in SI units.
We assume here that $ g_0, g_2 \geq 0 $. 
The functional derivative is given by
\begin{equation}
\frac{\delta U_n}{\delta n} =  g_0 n  -  g_2 \nabla^2 n  + \frac{\hbar^2}{8m} \left(
-\frac{2 \nabla^2 n }{n } + \frac{|\nabla n |^2}{n  ^2 }
\right).
\label{eq:functional_derivative}
\end{equation}
We want to minimize $ U_n $ with respect to a conservation constraint for $ n $. This can be achieved by minimizing 
\begin{equation}
\tilde{U}_n[n] =  U_n[n] - \mu \int \diff \vec{r} \left[
n \varone - \bar{n} 
\right]
,
\label{eq:tominimize}
\end{equation}
with respect to $ n $. Here $ \bar{n} $ is a given average particle number and $ \mu $ is the chemical potential. Extrema of $ \tilde{U}_n $ fulfill the condition
\begin{equation}
\frac{\delta U_n}{\delta n} = \mu
\end{equation}
i.e. the functional derivative is a constant.  Inserting a uniform $ n = \bar{n} $ in Eq.~\eqref{eq:functional_derivative} 
gives $  g_0 \bar{n} = \mu $ showing 
that $ n = \bar{n} $ is an extremum of $ \tilde{U}_n $. We will show that this exremum is a global minimum by showing that the energy   $ \tilde{U}_n $ is  \emph{globally convex}.

First some prerequisites \cite{boron1985variational}:
\begin{itemize}
	\item A functional is convex iff its second variation is non-negative for all test functions $ \varphi $. 
	\item The sum of convex functionals is convex.
\end{itemize}
The second variation is given by
\begin{equation}
\delta_{\varphi}^2 U_n [n] = \lim\limits_{\epsilon_1, \epsilon_2 \to 0} \frac{d^2}{d \epsilon_1 d \epsilon_2 } U_n[n + \epsilon_1 \varphi + \epsilon_2 \varphi ].
\label{eq:second_variation_def}
\end{equation}
The variations for the parts proportional to $ g_0 $ and $ g_2 $ can be calculated easily giving
\begin{equation}
\notag
g_0 \int \diff \vec{r } \left[
\varphi \varone ^2  \right], \quad 
g_2 \int  \diff \vec{r} \, |\nabla \varphi \varone |^2 ,
\end{equation}
respectively. Both are clearly non-negative implying that the corresponding parts in $ U $ are convex. 
For the last part we have 
\begin{equation}
\notag
\begin{split}
&\frac{|\nabla n  + (\epsilon_1+\epsilon_2) \nabla \varphi  |^2}{n  + (\epsilon_1+\epsilon_2) \varphi } = \\& 
|\nabla n  + (\epsilon_1+\epsilon_2) \nabla \varphi  |^2
\sum_{k=0}^\infty \frac{(\epsilon_1 + \epsilon_2 )^k \varphi  ^k}{n  ^{k+1} } \\
&= 
\text{lot} + \frac{2 \epsilon_1 \epsilon_2 [n  \nabla \varphi  - \varphi  \nabla n  ]^2 }{n  ^3} 
+ \text{hot},
\end{split}
\end{equation}
where `lot' stands for \emph{lower order terms} and `hot' is \emph{higher order terms}. This gives the second variation 
\begin{equation}
\frac{\hbar^2}{4m} \int \diff \vec{r} \left\lbrace
\frac{  [n  \nabla \varphi  - \varphi  \nabla n  ]^2 }{n  ^3} 
\right\rbrace \geq 0
\end{equation}
completing the proof. Setting $ g_2 = 0 $ proves the same for the zeroth order expansion. 

\section{Parameters for Rydberg-dressed BECs}\label{app:RB_parameters}
Here we will calculate the interaction parameters $ q_0 $, $ u_\text{r} $, and $ u_\text{e} $ for a model system for Rydberg-dressed BECs described in Ref.~\cite{Henkel2012}. The calculation presented here can be repeated for any radially symmetric interatomic potential $ u $ whose Fourier transform's smallest extremum is a minimum. 

We will use parameters defined in Fig.~2 of Ref.~\cite{Henkel2012}.
The interaction potential is 
\begin{equation}
u (r) = u_\text{sw} (r) + u_\text{RB} (r), 
\end{equation}
where $ u_\text{sw} (r) = g \delta (0) $ is the collision part due to $ S $-wave scattering and
\begin{equation}
u_\text{RB} (r) = \frac{C_6}{r^6 + R_c^6}
\end{equation}
is the long range interaction of two excited alkaline Rydberg atoms. $ C_6 $,  $ R_c $,  and $ g $ are system parameters. 

We calculate the Fourier Transform
\begin{equation}
\begin{split}
&\hat{u}(k) = \int_{0}^{\infty} \int_{0}^{2 \pi} \diff \theta \diff r \left[
r e^{i kr \cos (\theta)} u (r)  \right] \\
&= g + \int_{0}^{\infty} \diff r \left[
\frac{ \pi  r C_6 r J_0(k r)}{R_c^6+r^6} \right],
\end{split}
\end{equation}
where $ J_0 $ is the zeroth Bessel function of the first kind. The remaining integral can be calculated numerically or expressed in terms of the Meijer $ G $-function \cite{Olver:2010:NHM:1830479}  as
\begin{equation}
\begin{split}
&\hat{u}(k) =\\&	
\frac{ \pi}{3} C_6  R_c^{-4} G_{0,6}^{4,0}\left( 
\begin{array}{c} \left(
\frac{ R_c k}{6}
\right)^6 ;
0,\frac{1}{3},\frac{2}{3},\frac{2}{3},0,\frac{1}{3} \\
\end{array}
\right)+ g.
\end{split}
\end{equation}
In order to obtain the parameters we fit the polynomial
\begin{equation}
\hat{u}^\text{f} (k) = u_\text{r} + u_\text{e} \left(
k^2 - q_0^2
\right)^2
\end{equation}
to the solution $ \hat{u} $. There are many ways for performing the fitting. This problem has been studied in the context of approximating direct pair correlation functions in the classical density functional theory of freezing with simpler energy functionals \cite{Emmerich2012}. Here we will fit $ \hat{u}^\text{f} $ by matching $ q_0 $ to the smallest minimum $ k_\text{m} $ of $ \hat{u} $  and by ensuring that the energy of the uniform and the supersolid solutions will be correct i.e. $ \hat{u}^\text{f} (0) = \hat{u} (0) $ and $ \hat{u}^\text{f} (q_0) = \hat{u} (q_0) $. We find
\begin{equation}
\begin{split}
&q_0 R_c  = \tilde{q}_0  \approx 4.8202,\\
&\frac{ R_c^4}{C_6} \left(  u_\text{r} - g \right) = \frac{2 R_c^4}{C_6} \hat{u} (q_0)= \tilde{u}_r  , \\
&\frac{ (q_0 R_c)^4 u_\text{e}}{C_6}  = \frac{2 R_c^4}{  C_6} \left(  
\hat{u} (0) - \hat{u} (q_0)
\right) = \tilde{u}_e  .
\end{split}
\end{equation}
The approximate values for these parameters are
$ \tilde{u}_r  \approx -0.17023 $
$ \tilde{u}_e  \approx 3.9690 $.

We set $ g = 0 $ in accordance with Ref.~\cite{Henkel2012}. The $ \alpha_u $ parameter can be calculated as
\begin{equation}
\alpha_u = \frac{u_\text{r}}{q_0^4 u_\text{e}} = \frac{\hat{u} (q_0)}{\hat{u} (0) - \hat{u} (q_0)} = \frac{\tilde{u}_r}{\tilde{u}_e}
\end{equation}
giving $ \alpha_u \approx  -0.042889$.
We also need the average density. In Ref.~\cite{Henkel2012} the authors study a system of $ 10^4 $ Rb atoms in a volume of approximately $V \approx 9 R_c^2 $. We have
\begin{equation}
\begin{split}
\bar{\rho} &= \frac{m u_\text{e} q_0^2}{\hbar^2} 
\bar{n} = \frac{m }{R_c^2 \hbar^2} \frac{ \tilde{u}_e}{\tilde{q}_0^2} \frac{N}{V} C_6 \\ &
= \frac{m }{R_c^2 \hbar^2} \frac{ \tilde{u}_e}{\tilde{q}_0^2} \frac{N}{9 R_c^2} \frac{\hbar^4 c_6}{m^3 \omega_\text{tr}^2} \\&
= \frac{\hbar^2}{m^2 \omega_\text{tr}^2} \frac{1}{R_c^4}\frac{ \tilde{u}_e}{\tilde{q}_0^2} \frac{N c_6}{9}  \\&
= \left( \frac{l}{r_c} \right)^4 \frac{ \tilde{u}_e}{\tilde{q}_0^2} \frac{N c_6}{9} 
\approx 9.4.
\end{split}
\end{equation}
Here $ \omega_\text{tr}/2\pi = 125 $ Hz is the trapping frequency of a harmonic potential, $ l = \sqrt{\hbar/m \omega_\text{tr}}$ is the characteristic length scale for this trapping potential. The lower case characters are used for units defined by $ \omega_\text{tr} $ and a tilde is used for units defined by $ C_6 $ and $ R_c $. For parameters, see Tab.~\ref{tab:rydberg_parameters}.
\begin{table}
	\centering
	\caption{Parameter values for Rydberg-dressed BEC. }
	\begin{tabular}{l|l|r|l}
		Parameter & Alternative expr. & Value & Units \\
		\hline \hline
		$ l $ & $ \sqrt{\hbar/(m \omega_\text{tr})} $ & 0.972674 & $ \mu $m \\
		$ r_c $ & $  R_c/l $ & 2.65 & - \\
		$ c_6 $ & $ C_6 m^3 \omega^2/\hbar^4 $ & 2.45 & - \\
		$ N $ & - & $  10^4 $ & - \\
		$ q_0 $ & $ \tilde{q}_0/R_c $ & 1.87005 & $ \mu \text{m}^{-1}$ \\
		$ u_\text{e} $ & $ \tilde{u}_\text{e} C_6/(R_c^4 q_0^4) $ & 0.197182399 & $ \hbar^2/(m q_0^4) $ \\
		$ u_\text{r} $ & $  \tilde{u}_\text{r} C_6/(R_c^4) $ & $ -0.008456958$  & $ \hbar^2/m $ \\
	\end{tabular}
	\label{tab:rydberg_parameters}
\end{table}

The one-mode approximation predicts the phase transition at $ \alpha_u \approx -0.0245398 $ implying that these parameters should be well within the solid regime. The nearest neighbor distance for the hexagonal lattice is 
\begin{equation}
a_\text{nn} = \frac{4\pi}{q_0 \sqrt{3}},
\end{equation}
which for the parameters in \cite{Henkel2012} gives $ a_\text{nn} \approx 1.50516\, R_c = 3.87969\, \mu$m.

\begin{figure}[h]
	\centering
	\includegraphics[width = 3in]{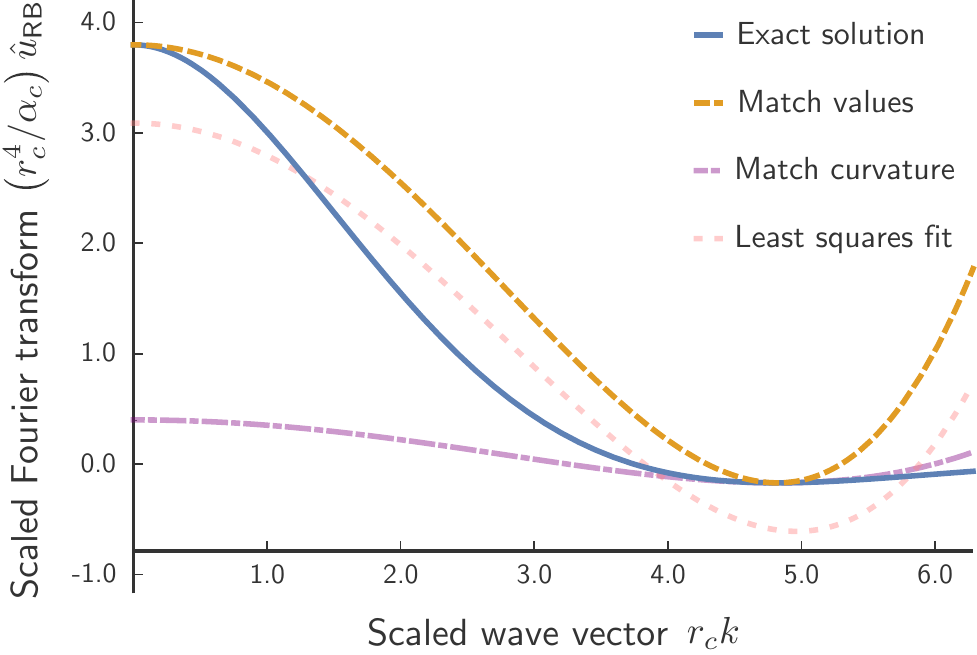}
	\caption{The Rydberg contribution of the Fourier transformed interaction potential $ \hat{u}_\text{RB} $ and different fitting polynomials. }
	\label{fig:rydberg_fit}
\end{figure}

Figure \ref{fig:rydberg_fit} shows the exact solution for $ \hat{u}_\text{RB} $ and several different fits. We use a fitting procedure matching the values of the fitted function with the exact solution at $ k=0 $ and $ k=q_0 $. We  also ensure that $ q_0 $ minimizes  $ \hat{u}_\text{RB} $. Other fits shown in Fig.~\ref{fig:rydberg_fit} include matching the curvature and the magnitude at $ k = k_m = q_0 $. This would most likely approximate elastic response of the solid phase better since it describes the energy well for modes $ k $ close to $ q_0 $. However, for this system this fitting method would give an inaccurate estimate for the energy of the uniform state described by $ \hat{u} (0) $. The last fit shows a general least squares fitting to $ \hat{u}_\text{RB} $. This fitting method fails to predict the wavelength of the one-mode approximation, the solid energy and the energy of the uniform state implying that global fitting might not work in this case. For more details see the Mathematica notebook \textsf{parameters.nb}.

%
%

\section{Numerical methods}\label{app:numerical_methods}

\subsection{Ground state determination}

The ground state of the system is found by time-evolving the non-conserved dissipative dynamical equation [Eq.~\eqref{eq:non-local_dissipative_dynamics}] except in the case of the quantum droplet phase for which we used overdamped dissipative wave dynamics [Eq.~\eqref{eq:dissipative_wave_formalism}]. In general the performance of the density formalism 
is better because the high order derivatives appear linearly in the dynamical equations. However, the density formalism becomes numerically unfeasible whenever the one-particle density $ \rho $ is close to zero due to difficulties in evaluating the quantum potential $ Q $. 

For the ground state calculations we impose doubly-periodic boundary conditions and calculate the derivatives in Fourier basis. We use a semi-implicit time-stepping algorithm that treats the linear parts of the equation implicitly and the non-linear parts explicitly \cite{Tegze2009}. To ensure numerical stability we require that the difference of the total energy $ \Delta E $ between time steps is non-positive. This difference is also used as a stopping condition: whenever $ \Delta E <  10^{-12}$ the iterative search for the ground state is stopped. 

The simulation domain is a rectangle of size $(4 \pi q^{-1} / \sqrt{3}, 4 \pi q^{-1})$.
For the periodic ground states the size of the domain has to be minimized in order to obtain the physical ground state. This is done as follows: Let some lowest mode of the ground state of a given domain be $ \exp (i \vec{q} \cdot \vec{x}) $. Now, $ q = |\vec{q}| $ defines the size of the domain. Due to the effect of the quantum potential the $ q $ that minimizes the ground state energy is usually a little bit less than 1. In order to find the minimal $ q $ we define an iterative process. We make an initial guess for $ q $ and calculate a number of points around it. This gives us a chart $ (q_n, E_n) $ to which we fit a parabola. We calculate analytically the minimum of this parabola and define a new set of $ q_n $ around this minimum and repeat the process. Let $ q_\text{m}^{(i)} $ be the minimal $ q $ calculated from the fitted parabola at $ i $th iteration. We stop the iteration when $ |q_\text{m}^{(i)} - q_\text{m}^{(i-1)}| $ is less than $ 10^{-5} $.

%
%

\subsection{Lattice vibration simulations}

Lattice vibrations are studied by simulating the full dynamics in the hydrodynamic formulation using the spectral PDE solver Dedalus \cite{dedalus}.
Since the mean field potential terms are linear in the density $\rho$, they can be integrated implicitly in the hydrodynamic formulation, eliminating the stiff timestepping restriction that would arise from integrating the higher-order potential terms explicitly.
Since spectral methods have no numerical dissipation, we regularize the equations by inserting a viscosity-like  term $\propto \nabla^2 \vec{v}$ in the velocity equation, which smooths small-scale variations and slowly damps velocity perturbations.

First, the ground state of the periodic domain is calculated after which
the vibration simulations are initialized with a small-amplitude plane-wave velocity signal.
The system was then evolved for $\sim 10^3$ temporal units.
The total kinetic energy of the system is observed to undergo oscillations which decay due to the added viscous term.
The square of an exponentially decaying sinusoid is fit to the time series of the kinetic energy to determine the oscillation frequency of the velocity signal.
The imposed perturbation wavelength and this fit frequency are used to assess the agreement between the simulations and the analytical dispersion relation.

\bibliography{supersolids}

\end{document}